\documentclass[a4paper,11pt]{article}
\pdfoutput=1 % if your are submitting a pdflatex (i.e. if you have
\usepackage{jcappub} % for details on the use of the package, please
\usepackage{graphicx}% Include figure files
\usepackage{dcolumn}% Align table columns on decimal point
\usepackage{bm}
\usepackage{amsmath,amsfonts,amssymb}
\usepackage{graphicx,epsfig}
\usepackage{color}
\usepackage{captcont}
\usepackage{slashed}
\usepackage{epstopdf}
\usepackage{mathtools}
\usepackage{natbib}
\usepackage{ulem}
\usepackage{subfigure}
\usepackage{graphicx}
\usepackage{CJKutf8}                   % see the JCAP-author-manual

\usepackage[T1]{fontenc} % if needed

\title{\boldmath Influence of quantum correction on the Schwarzschild black hole polarized image}

\author[a,b]{Sen Guo,\note{Corresponding author.}}
\author[c]{Yu-Xiang Huang}
\author[b]{Kuan Liu}
\author[b]{En-Wei Liang}
\author[d]{Kai Lin}

\affiliation[a]{College of Physics and Electronic Engineering, Chongqing Normal University,\\Chongqing 401331, People's Republic of China}
\affiliation[b]{Guangxi Key Laboratory for Relativistic Astrophysics, School of Physical Science and Technology, Guangxi University,\\Nanning 530004, People's Republic of China}
\affiliation[c]{School of Physics and Astronomy, China West Normal University,\\Nanchong 637000, People's Republic of China}
\affiliation[d]{Hubei Subsurface Multi-scale Imaging Key Laboratory, School of Geophysics and Geomatics, China University of Geosciences,\\Wuhan 430074, People's Republic of China}

\emailAdd{sguophys@126.com}
\emailAdd{yxhuangphys@126.com}
\emailAdd{liuk@st.gxu.edu.cn}
\emailAdd{lew@gxu.edu.cn}
\emailAdd{lk314159@hotmail.com}

\abstract{Using a model of an accretion disk around a Schwarzschild black hole, the analytic estimates for image polarization were derived by Narayan $et~al.$. [Astrophys. J, 102, 912 (2021)]. Recently, the EHT team also obtained polarization images of the Sgr A$^{*}$ and measured both linear and circular polarization [Astrophys. J. Lett, 964, L25 (2024)]. We find that quantum correction effects can also influence polarization information. Considering the quantum corrected Schwarzschild black hole (Kazakov-Solodukhin black hole), we derive the polarization intensity of the target black hole and investigate polarization images under different parameters. It is found that a larger quantum deformation leads to an expansion of the polarization region, while the polarization intensity value decrease. Under different observation angles, magnetic fields, fluid direction angles, and fluid velocity conditions, we also derive polarization images of corrected black holes. These key indicators not only affect the intensity of polarization but also the direction of polarization. We establish the relationship between polarization intensity and quantum correction deformation parameters, revealing a gradual decline in polarization intensity with reduced radius and an anti-polarization behavior induced by the progressive increase in deformation parameters at a constant radius. Our analysis may provide observational evidence for quantum effect of general relativity.}

\begin{document}
\maketitle
\flushbottom

\section{Introduction}
\label{sec:intro}
In recent years, the field of black hole astrophysics has experienced remarkable advancements, prominently demonstrated by the groundbreaking imagery unveiled by the Event Horizon Telescope (EHT). These results not only elucidate the spacetime configuration of central black holes but also encapsulate dynamic details concerning the arrangement of neighboring matter. These EHT images depict supermassive black holes residing at the cores of galaxies, exemplified by Messier (M) 87$^{*}$ \cite{1,2,3,4,5,6} and Sagittarius (Sgr) A$^{*}$ \cite{7,8,9,10,11,12}. These visual representations not only corroborate the predictions of general relativity (GR) but also pave the way for novel avenues of investigating the distribution of matter and electromagnetic interactions surrounding black holes. Prior to EHT observations, research on black holes predominantly centered on the classical framework of gravity theories, characterized by the Schwarzschild-like and Kerr-like black hole models \cite{Synge,Bardeen}. However, with our deepening understanding of extreme gravitational environments, classical models appear constrained in their capacity to elucidate specific observational phenomena. The existence of singularities, notably, engenders the incompleteness of geodesics, thereby affording researchers the opportunity to address this issue through the framework of quantum gravity theory. Recent theoretical and numerical simulation investigations, exemplified by loop quantum gravity theory, have provided a theoretical framework that transcends conventional categories \cite{13}. This offers a distinctive avenue for interpreting and elucidating observational data related to black holes through innovative theoretical models.

In 2021, the EHT collaboration unveiled polarized images of M87$^{*}$ \cite{14,15}. In the latest observations, The linear polarization images reveal a striking polarization pattern within the emission ring, showcasing a conspicuous spiral alignment of electric vectors. Notably, the western section of the ring exhibits a peak fractional polarization of approximately 40\% \cite{sgr-1,sgr-2}. In black hole astrophysics, polarization information in radiation serves as pivotal indicators of magnetic field configurations and the surrounding matter's behavior. To extract polarization information from black hole images observed in the sky, solving null geodesic equations of photon trajectories in the black hole spacetime. Moreover, parallel transport equations must be addressed to capture the evolution of polarization vectors along these photon geodesics. Recent studies have employed a simplified model to explore polarized images arising from synchrotron emission by axisymmetric fluids orbiting black holes within various magnetic field configurations \cite{16,17}. Synchrotron radiation from hot gas near a black hole generates a polarized image, with polarization intricately influenced by factors such as the magnetic field orientation in the emitting region, relativistic motion of the gas, the strong gravitational lensing effect of the BH, and parallel transport in the curved spacetime.

Using the Schwarzschild black hole model and considering an equatorial accretion disk around the black hole, Narayan $et~al.$ applied an approximate expression for null geodesics derived by Beloborodov \cite{18} and the conservation of the Walker-Penrose constant to provide analytical estimates for image polarization \cite{16}. This results offers comprehensive analytical approximations tailored specifically for Schwarzschild black holes. The simplistic model successfully replicates the electric vector position angle (EVPA) pattern and relative polarized intensity observed in the EHT images of M87$^{*}$. This comparison, made under the reasonable assumption that the magnetic field aligns with the fluid velocity, corresponds with the clockwise rotation inferred from total intensity images. Expanding upon the Schwarzschild black hole scenario, a toy model describing an equatorial source around a Kerr black hole in investigated, obtaining the radiation polarization images. The research considered the geometric effects of black hole spin on photon parallel transport, isolating these effects from the intricate interplay of gravity and electromagnetic processes in the emission region \cite{17}. These investigations unveil that the observed polarization characteristics in black hole images are predominantly shaped by magnetic field geometry, in conjunction with the black hole's intrinsic parameters and the observer inclination.

Apart from examining polarized images within Schwarzschild and Kerr spacetimes, the influence of various modified gravity on black hole polarization is investigated. Considering the interaction between photons and the Weyl tensor, the impact of Weyl tensor on the polarization image of Schwarzschild black holes is discussed, which showed a clear couping regions in the polarization image of the Schwarzschild black hole \cite{19}. Investigating the polarized image arising from an equatorial emitting ring encircling a four-dimensional Gauss-Bonnet black hole, it demonstrated that the influence of the Gauss-Bonnet parameter on the polarized image is contingent upon the magnetic field configuration, observer inclination angle, and fluid velocity \cite{20}. Palumbo $et~al.$ decomposed images depicting half-orbits of photons within the accretion environment of M87$^{*}$ sourced from the simulated image repository of the EHT, whose find a relative depolarization within the photon ring in magnetically arrested disk simulation images, and the result display that an approximate 50$\%$ depolarization within the photon ring region in comparison to the direct image \cite{21}. Similar features are also discerned across various black hole scenarios, encompassing polarized images of synchrotron radiation emanating from electron sources within Schwarzschild black hole spacetime with a vertical and uniform magnetic field \cite{22}, regular-Bardeen and Hayward black holes \cite{23}, Bonnor black holes \cite{24}, traversable wormholes \cite{25}, rotating black hole within Scalar-Tensor-Vector-Gravity theory \cite{26}, Kerr black holes within weak magnetic field \cite{27}, rotating black holes encircled by a cold dark matter halo \cite{28}, and naked singularities \cite{29}.

\par
In GR, the interplay between the metric and matter fields is governed by the Einstein field equations. When considering spherically symmetric quantum fluctuations, it is customary to examine the potential ramifications of quantum effects on the Schwarzschild metric. Within this context, quantum fluctuations may introduce corrections to the Schwarzschild metric, suggesting potential modifications at small scales. The quantum corrected black hole is derived in Ref. \cite{30}. Subsequent investigations have probed quantum phenomena in this black hole, event horizons, and spherically symmetric gravitational collapses, indicating that quantum effects could hinder the formation of inherent singularities at the origin \cite{31,32,33}. Kazakov $et~al.$ explored the deformation of the Schwarzschild metric induced by quantum fluctuations, evolving the four-dimensional gravity theory with Einstein action into an effective two-dimensional dilaton gravity theory. Expanding upon this notion, Kazakov and Solodukhin derived the quantum corrected Schwarzschild solution \cite{34}. This series of studies is motivated by the disruptive nature of singularities on the complete relationship of geodesics within classical GR, rendering its study ineffective. Hence, there arises a necessity to incorporate quantum effects surrounding singularities.

On the other hand, quantum corrected black holes exhibit discrepancies from classical black holes concerning the photon sphere and shadows, primarily due to the significant quantum effects in the strong gravitational region. The investigation of quasinormal modes delves into the impact of quantum deformation on the radius of the black hole's shadow, underscoring the pivotal role of quantum corrections in the observation of black hole shadows \cite{35}. Furthermore, the examination of weak and strong deflection gravitational lensing by quantum deformation and its observable effects is detailed \cite{36}. In a study focusing on an optically and geometrically thin disk, Peng $et~al.$ scrutinized the shadow of a quantum correction Schwarzschild black hole, revealing observable alterations in the appearance of the accretion disk around the black hole \cite{37}. Building upon this research, we have previously explored the influence of quantum correction on the optical appearance of black hole accretion disks, demonstrating a gradual decrease in the radiation flux with increasing quantum deformation \cite{38}. In the work previously undertaken by our team, utilizing observational data from the IXPE telescope, we have discovered that the X-ray radiation from the Crab pulsar wind nebula exhibits a global polarization degree of 45\%, exceeding a confidence level of over 30 standard errors \cite{Na}. Moreover, we have found localized regions with polarization degrees as high as 63\%. These observations approach the theoretical limit predicted by the theory of synchrotron radiation from electrons in a completely ordered magnetic field, revealing an extremely ordered magnetic field within the pulsar wind nebula. This suggests that its X-ray radiation originates from synchrotron radiation by electrons, imposing strict constraints on models of particle acceleration mechanisms. Motivated by these investigations, we delve into the influence of quantum correction on the polarization image of the black hole in this analysis. This article serves as a continuation of prior work, aiming to achieve a more comprehensive understanding of the effects of quantum correction on the appearance of black holes.

We will conduct a comprehensive analysis of black hole polarized images within the framework of an equatorial synchrotron emitting ring, using the Kazakov-Solodukhin (KS) black hole as a case study. The structure of this paper is as follows: In Section \ref{sec:2}, we provide a review of the KS black hole and employ the Walker-Penrose method to compute the polarized images observed in the orbital fluid model. Section \ref{sec:3} is dedicated to deriving the polarization image of the KS black hole. In Section \ref{sec:4}, we showcase the polarization image of the KS black hole juxtaposed against the background of an accretion disk. Additionally, we juxtapose the polarization characteristics of this black hole with those of M87$^{*}$. Finally, we conclude by summarizing the key findings of the entire paper.

\section{Observed polarization field in quantum corrected Schwarzschild black hole}
\label{sec:2}
This portion of the text focuses on the spacetime associated with the KS black hole. The metric describing the static spherically symmetric black hole can be expressed as follows:
\begin{equation}
\label{2-1}
{\rm d}s^{2}=-A(r){\rm d}t^{2}+B(r){\rm d}r^{2}+C(r)({\rm d}\theta^{2}+\sin^{2}\theta {\rm d}\phi^{2}).
\end{equation}
The function $A(r)$ represents the metric potential of the black hole, with its reciprocal denoted as $B(r)$, while $C(r)$ is defined as $r^2$. Kazakov and Solodukhin discussed the deformation of the Schwarzschild black hole caused by spherically symmetric quantum fluctuations, the metric potential of the KS black hole is given by \cite{34}
\begin{equation}
\label{2-2}
A(r)=\frac{\sqrt{r^{2}-a^{2}}}{r}-\frac{2M}{r},
\end{equation}
where $M$ represents the black hole mass, while the parameter $a$ serves as a deformation factor in the context of effective two-dimensional dilaton gravity. The Schwarzschild black hole corresponds to a simplified form of Eq. (\ref{2-2}) in the limit where $a$ equals zero. The horizon radius of the KS black hole can be represented as:
\begin{equation}
\label{2-3}
r_{\rm H}=\sqrt{a^{2}+4M^{2}}.
\end{equation}

\par
In accordance with references \cite{16,17}, we investigate a physical scenario involving a slender synchrotron-emitting ring in orbital motion within the equatorial plane of a KS black hole. This ring rotates at a consistent coordinate radius amidst a constant local magnetic field, resulting in the emission of linearly polarized synchrotron radiation. The distant observer adopts a face-on orientation, deviating angularly by $\theta_{0}$ from the normal direction of the emitting ring. A fluid element, denoted as $P$ and positioned at an azimuthal angle $\phi$ measured from the line of nodes connecting the ring plane and the observer's sky plane. We establish Cartesian coordinates within the geodesic plane, aligning the unit vector along the x-axis, $\hat{x}$, with $OP$, where the observer resides in the $\hat{x}-\hat{z}$ plane. The observable polarization at significant distances is computed by parallel-transporting the polarization vector along null geodesics and determining its projection. The relationship between the angle $\psi$ and the azimuthal angle $\phi$ is given by:
\begin{equation}
\label{2-4}
\psi=-\arccos\big(-\sin\theta_{0} \sin\phi\big).
\end{equation}

\par
Investigating a null geodesic trajectory originating from point $P$ and reaching the observer, where the conserved energy is denoted as $k_{\rm t}=1$, the orthogonal temporal and spatial components of the photon's momentum at the location $P$ in the G-frame can be formulated as:
\begin{align}
\label{2-5}
k_{\rm G}^{\hat{t}}&=-\frac{k_{\rm t}}{\sqrt{\frac{\sqrt{R^{2}-a^{2}}}{R}-\frac{2M}{R}}},\nonumber\\
k_{\rm G}^{\hat{x}}&=k_{\rm G}^{\hat{t}}\cos\alpha,~~k_{\rm G}^{\hat{y}}=0,~~k_{\rm G}^{\hat{z}}=k_{\rm G}^{\hat{t}}\sin\alpha.
\end{align}

\par
Defining $\alpha \equiv \arccos\Big(\cos\psi+\frac{2}{R}(1-\cos\psi)\Big)$, we shift to a Cartesian frame synchronized with the motion of the fluid ring. This transition between frames entails a rotation of an angle $\xi$ around the $\hat{x}$ axis. Notably, both the G-frame and P-frame share a common $\hat{x}$ axis. By rotating by an angle $\xi$, interchangeability between these reference frames is attained. Hence, the formal representation of this angle is given by \cite{16,17}:
\begin{equation}
\label{2-6}
\xi=\arcsin\Bigg(\frac{\sin\theta_{0}\cos\phi}{\sin\psi}\Bigg).
\end{equation}
Applying this rotation to the orthonormal components of $k_{\rm G}$ yields the corresponding orthonormal components within the P-frame, we have
\begin{align}
\label{2-7}
&k_{\rm P}^{\hat{t}}=\frac{1}{\sqrt{\frac{\sqrt{R^{2}-a^{2}}}{R}-\frac{2M}{R}}},~~~~~k_{\rm P}^{\hat{x}}=\frac{\cos\alpha}{\sqrt{\frac{\sqrt{R^{2}-a^{2}}}{R}-\frac{2M}{R}}},\nonumber\\
&k_{\rm P}^{\hat{y}}=-\frac{\sin\xi\sin\alpha}{\sqrt{\frac{\sqrt{R^{2}-a^{2}}}{R}-\frac{2M}{R}}},~~~k_{\rm P}^{\hat{z}}=\frac{\cos\xi\sin\alpha}{\sqrt{\frac{\sqrt{R^{2}-a^{2}}}{R}-\frac{2M}{R}}}.
\end{align}
Assuming the fluid moves with a velocity denoted as $\overrightarrow{\beta}$ within the $(r)-(\phi)$ plane of the local frame, further parameterization of the fluid velocity can be attained as \cite{16,17}
\begin{equation}
\label{2-8}
\overrightarrow{\beta}=\beta\cos\chi(r)+\beta\sin\chi(\phi).
\end{equation}
Here, $\beta$ represents the magnitude of the flow velocity, and $\chi$ denotes the angle of fluid direction. For M87$^{*}$, the rotation occurs clockwise, indicating $\sin\chi<0$. It is crucial to stress that $\overrightarrow{\beta}$ solely characterizes fluid motion and remains uninfluenced by the shape of the ring. The orthogonal components of $k_{\rm F}^{\hat{t}}$ in the fluid frame (F-frame) can be determined through a Lorentz boost, as outlined in \cite{16,17}:
\begin{align}
\label{2-9}
k_{\rm F}^{\hat{t}}=&\gamma k_{\rm P}^{\hat{t}} - \gamma \beta \cos\chi k_{\rm P}^{\hat{x}} - \gamma \beta \sin\chi k_{\rm P}^{\hat{y}},\nonumber\\
k_{\rm F}^{\hat{x}}=&- \gamma \beta \cos\chi k_{\rm P}^{\hat{t}} + \big(1+(\gamma-1)\cos^{2}\chi\big)k_{\rm P}^{\hat{x}} \nonumber\\
                &+ (\gamma-1)\cos\chi \sin\chi k_{\rm P}^{\hat{y}},\nonumber\\
k_{\rm F}^{\hat{y}}=&- \gamma \beta \sin\chi k_{\rm P}^{\hat{t}} + \big(1+(\gamma-1)\sin^{2}\chi \big)k_{\rm P}^{\hat{y}} \nonumber\\
                &+ (\gamma-1)\cos\chi \sin\chi k_{\rm P}^{\hat{x}},\nonumber\\
k_{\rm F}^{\hat{z}}=&k_{\rm P}^{\hat{z}}.
\end{align}
The Lorentz factor is denoted as $\gamma$, defined by $\gamma=\frac{1}{\sqrt{1-\beta^{2}}}$. Any radiation emitted along $k_{\rm F}^{\hat{u}}$ in the F-frame undergoes a Doppler shift by the time it reaches the observer. This Doppler factor encompasses both gravitational redshift and Doppler shift resulting from velocity, expressed as \cite{16,17}:
\begin{equation}
\label{2-10}
\delta=\frac{1}{k_{\rm F}^{\hat{t}}}.
\end{equation}

\par
We consider the fluid situated within a constant local magnetic field, and we represent its components in the rest frame as outlined in \cite{16,17}:
\begin{equation}
\label{2-11}
\overrightarrow{B}=B_{\rm r}\hat{x} + B_{\rm \phi}\hat{y} + B_{\rm z}\hat{z}.
\end{equation}
The angle $\zeta$ between the magnetic field $\overrightarrow{B}$ and the 3-vector $\overrightarrow{k}_{F}$ satisfies \cite{16,17}:
\begin{equation}
\label{2-12}
\sin\zeta = \frac{|\overrightarrow{k}_{\rm F} \times \overrightarrow{B}|}{|\overrightarrow{k}_{\rm F}||\overrightarrow{B}|}.
\end{equation}
In the fluid frame, the E-vector of the radiation aligns with $\overrightarrow{k}_{\rm F} \times \overrightarrow{B}$. Consequently, we express the orthonormal components of the polarization 4-vector $f^{\rm \mu}$ as detailed in \cite{16,17}:
\begin{align}
\label{2-13}
&f_{\rm F}^{\hat{t}}=0,~~~~~~~~~~~~~~~~~~~~~f_{\rm F}^{\hat{x}}=\frac{\big(\overrightarrow{k}_{\rm F} \times \overrightarrow{B}\big)_{\hat{x}}}{|\overrightarrow{k}_{\rm F}|},\nonumber\\
&f_{\rm F}^{\hat{y}}=\frac{\big(\overrightarrow{k}_{\rm F} \times \overrightarrow{B}\big)_{\hat{y}}}{|\overrightarrow{k}_{\rm F}|},~~~~~~f_{z}^{\hat{z}}=\frac{\big(\overrightarrow{k}_{\rm F} \times \overrightarrow{B}\big)_{\hat{z}}}{|\overrightarrow{k}_{\rm F}|}.
\end{align}
Here, the normalized polarization vector satisfies $f^{\rm \mu}f_{\rm \mu}=\sin^{2}\zeta|\overrightarrow{B}_{\rm F}|^{2}$ and $f^{\rm \mu}k_{\rm \mu}=0$. Upon applying the inverse Lorentz transformation, we retrieve the polarization 4-vector $f_{\rm F}^{\rm \mu}$ in the P-frame, resulting in \cite{16,17}:
\begin{align}
\label{2-14}
f_{\rm P}^{\hat{t}}=&\gamma f_{\rm F}^{\hat{t}} + \gamma \beta \cos\chi f_{\rm F}^{\hat{x}} + \gamma \beta \sin\chi f_{\rm F}^{\hat{y}},\nonumber\\
f_{\rm P}^{\hat{x}}=&\gamma \beta \cos\chi f_{\rm F}^{\hat{t}} + \big(1+(\gamma-1)\cos^{2}\chi \big)f_{\rm F}^{\hat{x}} \nonumber\\
                &+ (\gamma-1)\cos\chi \sin\chi f_{\rm F}^{\hat{y}},\nonumber\\
f_{\rm P}^{\hat{y}}=&\gamma \beta \sin\chi f_{\rm F}^{\hat{t}} + \big(1+(\gamma-1)\sin^{2}\chi \big)f_{\rm F}^{\hat{y}} \nonumber\\
                &+ (\gamma-1)\cos\chi \sin\chi f_{\rm F}^{\hat{x}},\nonumber\\
f_{\rm P}^{\hat{z}}=&f_{\rm F}^{\hat{z}}.
\end{align}
Since the Cartesian unit vectors $\hat{x}$, $\hat{y}$, $\hat{z}$ in the P-frame align with the spherical polar unit vectors $\hat{r}$, $\hat{\phi}$, $\hat{\theta_{0}}$ in the Schwarzschild black hole frame, the orthonormal components of $k$ and $f$ in the coordinates of the KS black hole, which is static and spherically symmetric, can be expressed as:
\begin{align}
\label{2-15}
k^{\rm t}&=\frac{1}{\sqrt{\frac{\sqrt{R^{2}-a^{2}}}{R}-\frac{2M}{R}}},~~~~k^{\rm x}=\sqrt{\frac{\sqrt{R^{2}-a^{2}}}{R}-\frac{2M}{R}}k_{\rm P}^{\hat{x}},\nonumber\\
k^{\rm y}&=-\frac{k_{\rm P}^{\hat{y}}}{R},~~~~~~~~~~~~~~~~~~~k^{\rm z}=\frac{k_{\rm P}^{\hat{z}}}{R}.
\end{align}
\begin{align}
\label{2-16}
f^{\rm t}=&\frac{f_{\rm P}^{\hat{t}}}{\sqrt{\frac{\sqrt{R^{2}-a^{2}}}{R}-\frac{2M}{R}}},~~~~f^{\rm x}=\sqrt{\frac{\sqrt{R^{2}-a^{2}}}{R}-\frac{2M}{R}}f_{\rm P}^{\hat{x}},\nonumber\\
f^{\rm y}=&-\frac{f_{\rm P}^{\hat{y}}}{R},~~~~~~~~~~~~~~~~~~f^{\rm z}=\frac{f_{\rm P}^{\hat{z}}}{R}.
\end{align}
In the spacetime of the KS black hole, the celestial coordinates $(x,y)$ corresponding to the path of a photon moving from point $P$ along the null geodesic towards the observer at infinity are delineated in \cite{39}:
\begin{align}
\label{2-17}
&x=-\frac{k_{\rm \phi}}{\sin\theta_{0}}=-\frac{Rk^{\hat{\phi}}}{\sin\theta_{0}},\\
&y=k_{\rm \theta}=R\Big[\big(k^{\hat{\theta}}\big)^{2}-\cot^{2}\theta_{0}\big(k^{\hat{\phi}}\big)^{2}\Big]^{\frac{1}{2}}sgn(\sin\phi).
\end{align}
In determining the polarization vector at the observer, we employ the Walker-Penrose constant denoted as $K_{1}+iK_{2}$ \cite{40}. The computation of the polarized vector at the observer becomes straightforward as both the real and imaginary parts of $K$ remain conserved along the null geodesic, which is
\begin{align}
\label{2-18}
&K_{1}=\Psi_{2}^{-1/3}(k^{\rm t}f^{\rm r}-k^{\rm r}f^{\rm t}),\nonumber\\
&K_{2}=-\Psi_{2}^{-1/3}R^2(k^{\rm \phi}f^{\rm \theta}-k^{\rm \theta}f^{\rm \phi}),
\end{align}
where $\Psi_{2}$ represents the second Weyl scalar. Within the spacetimes of the KS black hole, characterized by:
\begin{align}
\label{2-19}
\Psi_{2}=-\frac{2M+\frac{a^2}{\sqrt{R^2-a^2}}}{2R^{3}}.
\end{align}
We can evaluate the two transverse components of the electric field of polarization at the observer, denoted as $\overrightarrow{E}$, leading to \cite{41}:
\begin{align}
\label{2-20}
&E_{\rm x,norm}=\frac{yK_{2}+xK_{1}}{\sqrt{(K_{1}^{2}+K_{2}^{2})(x^{2}+y^{2})}},\nonumber\\
&E_{\rm y,norm}=\frac{yK_{1}-xK_{2}}{\sqrt{(K_{1}^{2}+K_{2}^{2})(x^{2}+y^{2})}},\nonumber\\
&E_{\rm x,norm}^{2}+E_{\rm y,norm}^{2}=1.
\end{align}

\par
In the context of synchrotron radiation, we can approximate the intensity of linearly polarized light reaching the observer from the source point $P$ as:
\begin{align}
\label{2-21}
|I|=\delta^{3+\alpha_{\rm v}} l_{\rm P} |\overrightarrow{B}|^{1+\alpha_{\rm v}} \sin^{1+\alpha_{\rm v}}\zeta.
\end{align}
Here the power law index $\alpha_{\rm v}$ is contingent upon the ratio of the emitted photon energy $h\nu$ to the temperature $kT$ of the disk. Denoting the geodesic path length taken by the photon within the emitting region as $l_{\rm P}$, it can be formulated as:
\begin{align}
\label{2-22}
l_{\rm P}=\frac{k_{\rm F}^{\hat{t}}}{k_{\rm F}^{\hat{z}}}H,
\end{align}
where the symbol $H$ denotes the height of the disk. Following references \cite{16,17}, when the parameter $\alpha_{\rm v}$ is set to 1, the components of the observed polarized vector along the $x$ and $y$ directions are expressed as:
\begin{align}
\label{2-23}
&E_{\rm x,obs}=\delta^{2} l_{\rm P} |\overrightarrow{B}| \sin\zeta E_{\rm x,norm},\nonumber\\
&E_{\rm y,obs}=\delta^{2} l_{\rm P} |\overrightarrow{B}| \sin\zeta E_{\rm y,norm},\nonumber\\
&|I|=E_{\rm x,obs}^{2}+E_{\rm y,obs}^{2}.
\end{align}
The total EVPA is
\begin{align}
\label{2-24}
EVPA=\arctan\Bigg(-\frac{E_{\rm x,obs}}{E_{\rm y,obs}}\Bigg)=\frac{1}{2}\arctan\frac{U}{Q},
\end{align}
where the $U$ and $Q$ are Stokes parameters, which is
\begin{align}
\label{2-25}
U=-2E_{\rm x,obs}E_{\rm y,obs},~~~~~~~Q=E^{2}_{\rm y,obs}-E^{2}_{\rm x,obs}.
\end{align}
Drawing from the preceding discussion, the influence of quantum correction on the black hole polarization image within an emitting ring context can be explored.

\section{Effect of quantum correction on black hole polarized image}
\label{sec:3}
In the preceding section, we employed an analytical model to characterize the synchrotron radiation emitted by a magnetized fluid ring. This model serves as the basis for simulating the observable polarization of a KS black hole. The computation of observable polarization at each point on the image is conducted following Eq. (\ref{2-23}). Here, the polarization intensity is directly linked to the length of the polarization vector, while deviations in the EVPA are reflected in the rotation of the vector around the ring.

\par
\begin{figure*}[htbp]
\centering
\includegraphics[width=4.8cm,height=4.8cm]{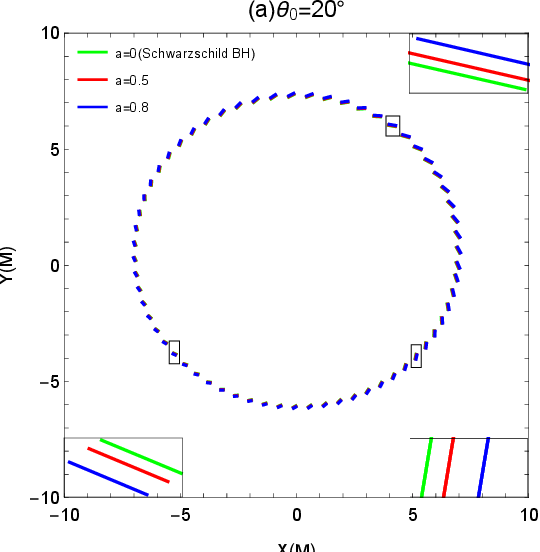}
\hspace{0.1cm}
\includegraphics[width=4.8cm,height=4.8cm]{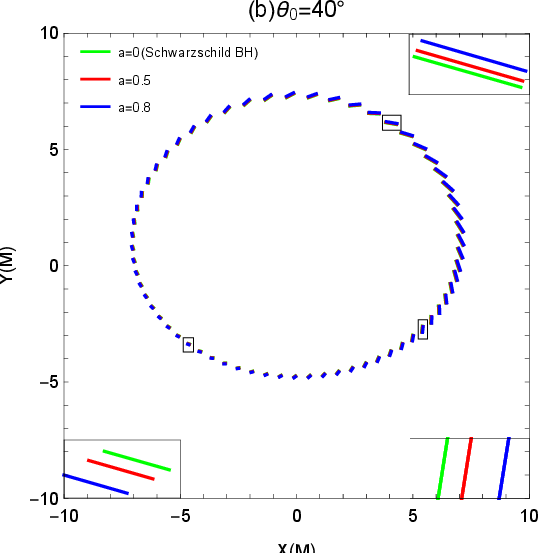}
\hspace{0.1cm}
\includegraphics[width=4.8cm,height=4.8cm]{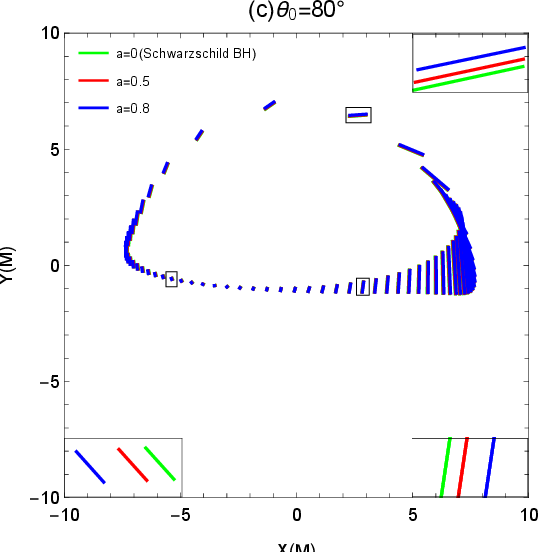}
\hspace{0.1cm}
\caption{The polarization intensity tick plots in the KS black hole spacetime under various values of the deformation parameter and observation angles. {\em Left Panel}: Observation angle $\theta_{0}=20^{\circ}$. {\em Middle Panel}: Observation angle $\theta_{0}=40^{\circ}$. {\em Right Panel}: Observation angle $\theta_{0}=80^{\circ}$. The black hole mass is fixed at $M = 1$, with an emitting ring radius of $R=6$. The fluid parameters comprise a velocity of $\beta=0.4$, a fluid direction angle of $\chi=-150^{\circ}$, and a magnetic field characterized by $B_{r}=0.87$, $B_{\phi}=0.5$, $B_{z}=0$.}
\label{fig:1}
\end{figure*}

The determination of observable polarization relies on numerous physical parameters inherent to the accreting flow, encompassing fluid velocity and magnetic field strength in the fluid's rest frame. Furthermore, the resulting image is shaped by spacetime parameters, including the spacetime metric, the position of the fluid ring, the observer inclination angle, and other pertinent factors. The initial step involves examining polarization images of KS black holes from various observation angles. We vary deformation parameters to determine the impact of quantum correction effects on polarization image. Drawing from the findings reported by the EHT, where the magnetic field configuration expressed as $B_{(\rm r,\phi,z)}=(0.87,0.5,0)$ is deemed optimal for elucidating polarization phenomena in M87$^{*}$, we adopt identical magnetic field parameters to simulate the observable polarization emanating from a stationary fluid ring. As illustrated in Fig. \ref{fig:1}, an increase in the observation inclination angle leads to a gradual flattening of the polarization image, heightened polarization intensity, and a progressive shift in polarization direction towards the southeast, reminiscent of the behavior observed in Schwarzschild black holes. Concurrently, augmented deformation parameters result in the expansion of the polarization region. Notably, regardless of the direction, the scenario with $a=0.8$ consistently manifests in the outermost layer, suggesting that the quantum correction effect contributes to enlarging the polarization region.
\begin{figure*}[ht]
\centering
\includegraphics[width=5cm,height=5cm]{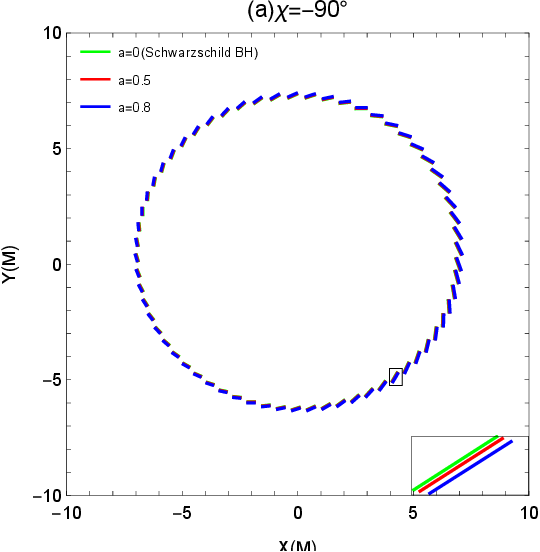}
\hspace{0.1cm}
\includegraphics[width=5cm,height=5cm]{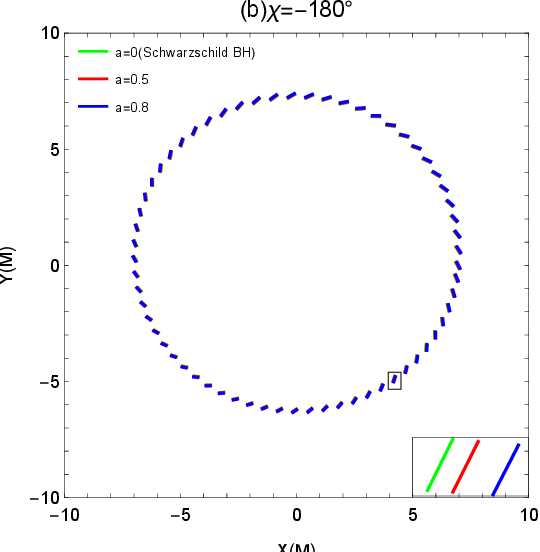}
\hspace{0.1cm}
\includegraphics[width=5cm,height=5cm]{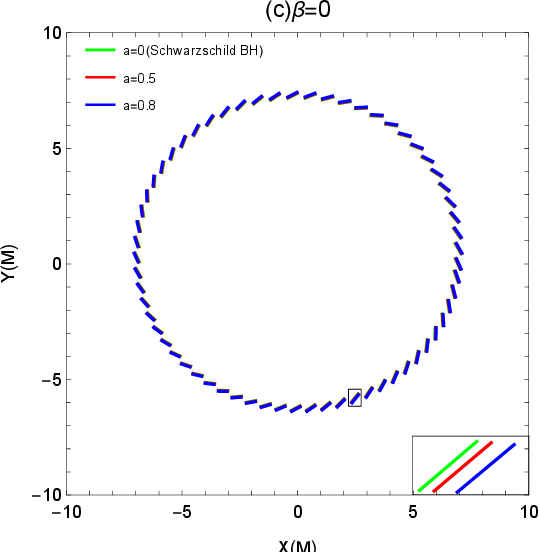}
\hspace{0.1cm}
\includegraphics[width=5cm,height=5cm]{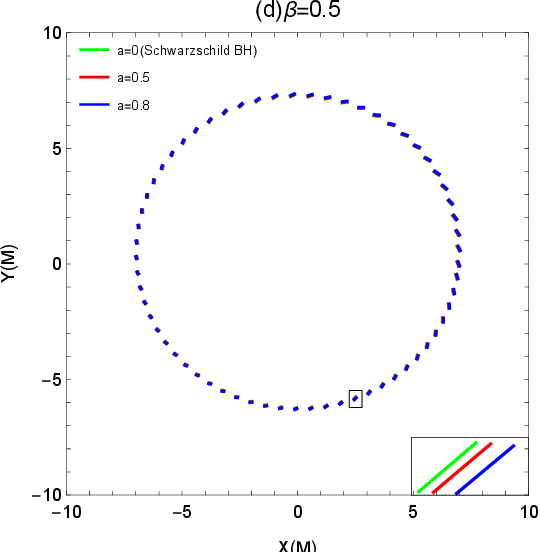}
\caption{The polarization intensity tick plots in the KS black hole spacetime under different values of the fluid direction angle and the fluid velocity. {\em (a) Panel}: Fluid direction angle $\chi=-90^{\circ}$. {\em (b) Panel}: Fluid direction angle $\chi=-180^{\circ}$. The fluid velocity of $\beta=0.3$. {\em (c) Panel}: Fluid velocity $\beta=0$. {\em (d) Panel}: Fluid velocity $\beta=0.5$. The fluid direction angle $\chi=-150^{\circ}$. Key parameters include a black hole mass of $M=1$, an emitting ring radius of $R=6$, an observation angle of $\theta_{0}=17^{\circ}$, and a magnetic field configuration of $B_{r}=0.87$, $B_{\phi}=0.5$, and $B_{z}=0$.}
\label{fig:2}
\end{figure*}

\par
In Fig. \ref{fig:2}, we depict the influence of fluid direction angle and fluid velocity on polarization images of KS black holes across various deformation parameters. Notably, transitioning from $\chi=-90^{\circ}$ to $\chi=-180^{\circ}$ induces a decrease in polarization intensity, accompanied by a slight increase in the slope of the EVPA. This observation underscores the combined impact of fluid direction on both polarization intensity and direction. Furthermore, an increase in fluid velocity corresponds to a reduction in polarization intensity without altering the potential angle of the electric vector. It's indeed noteworthy that despite variations in fluid direction angle and velocity, the influence of quantum correction persists with consistent properties. This suggests a spatiotemporal feature that remains independent of external conditions.
\begin{figure*}[htbp]
\centering
\includegraphics[width=2.5cm,height=2.5cm]{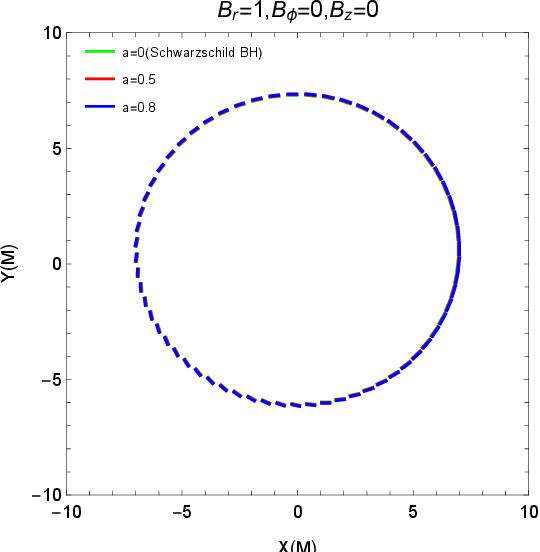}
\includegraphics[width=2.5cm,height=2.5cm]{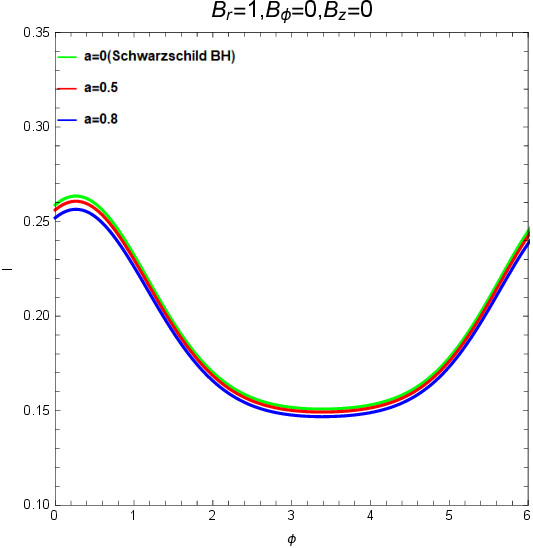}
\includegraphics[width=2.5cm,height=2.5cm]{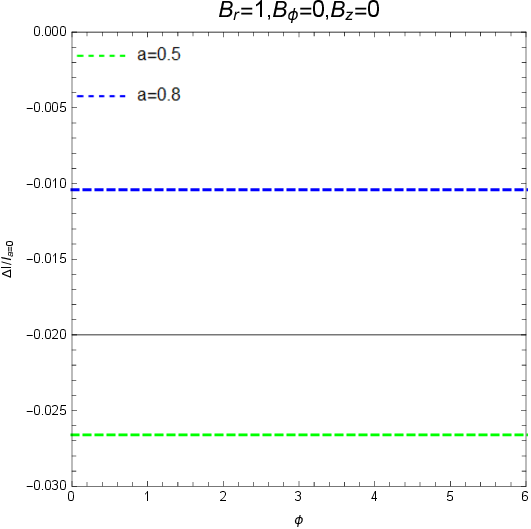}
\includegraphics[width=2.5cm,height=2.5cm]{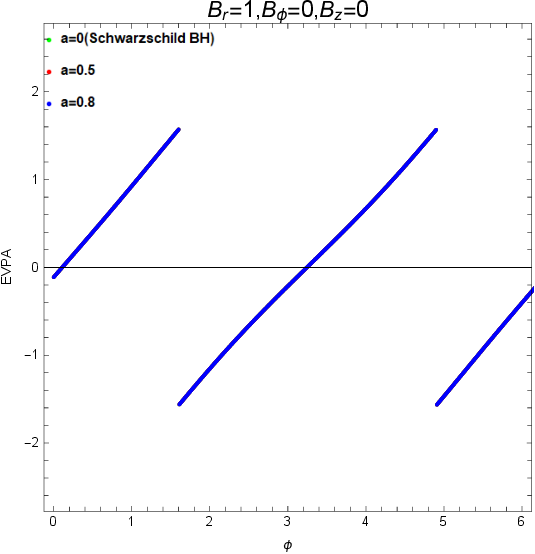}
\includegraphics[width=2.5cm,height=2.5cm]{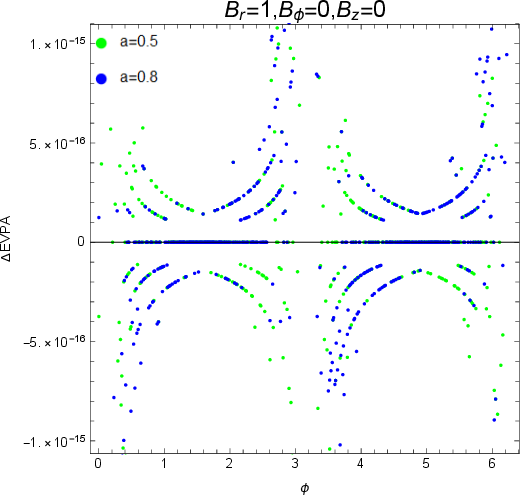}
\includegraphics[width=2.5cm,height=2.5cm]{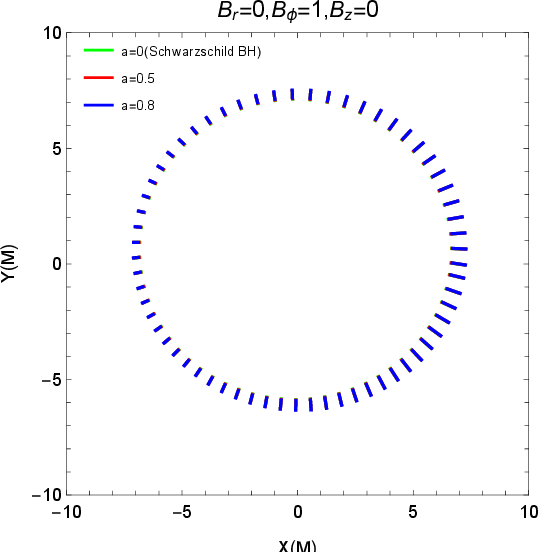}
\includegraphics[width=2.5cm,height=2.5cm]{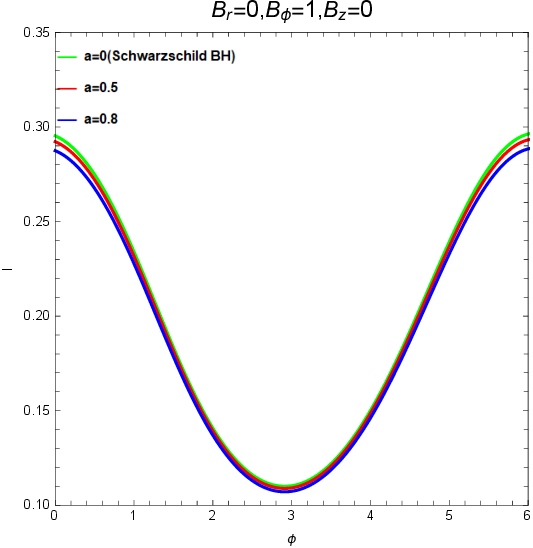}
\includegraphics[width=2.5cm,height=2.5cm]{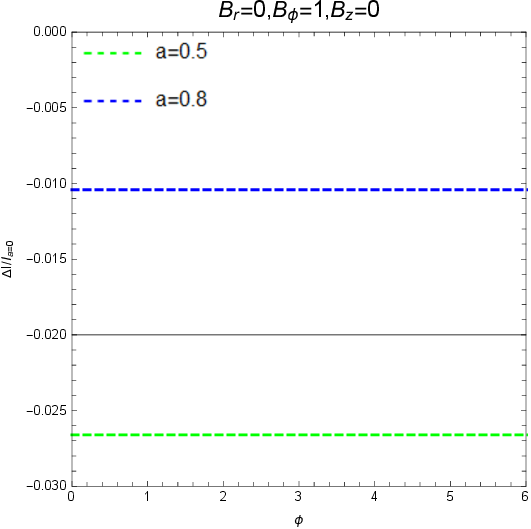}
\includegraphics[width=2.5cm,height=2.5cm]{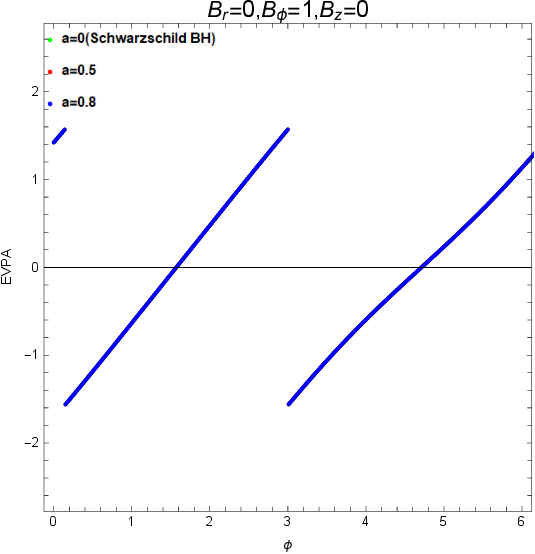}
\includegraphics[width=2.5cm,height=2.5cm]{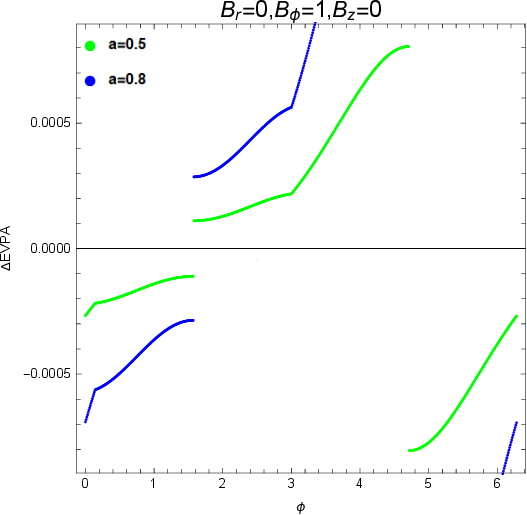}
\includegraphics[width=2.5cm,height=2.5cm]{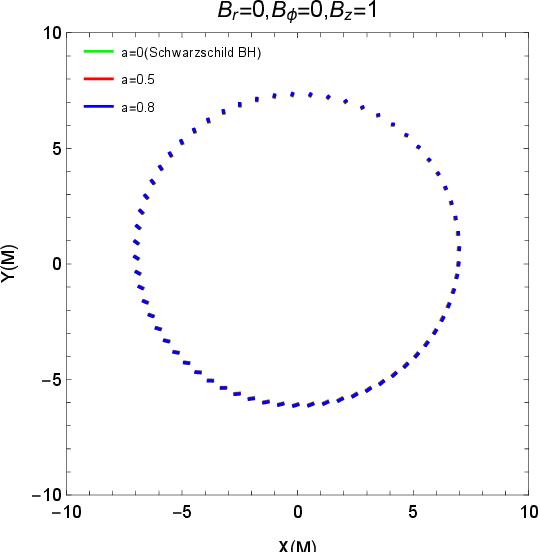}
\includegraphics[width=2.5cm,height=2.5cm]{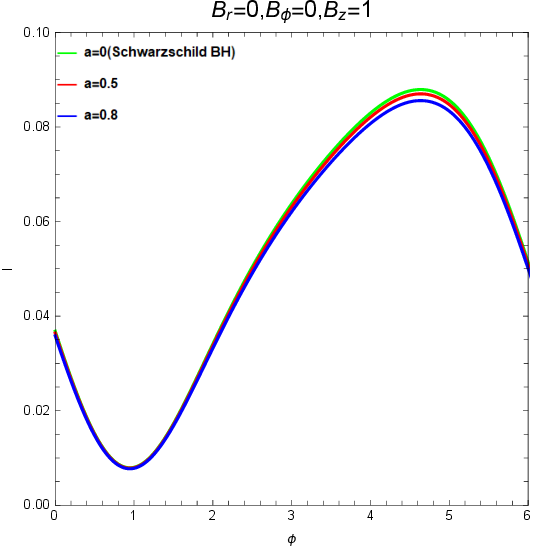}
\includegraphics[width=2.5cm,height=2.5cm]{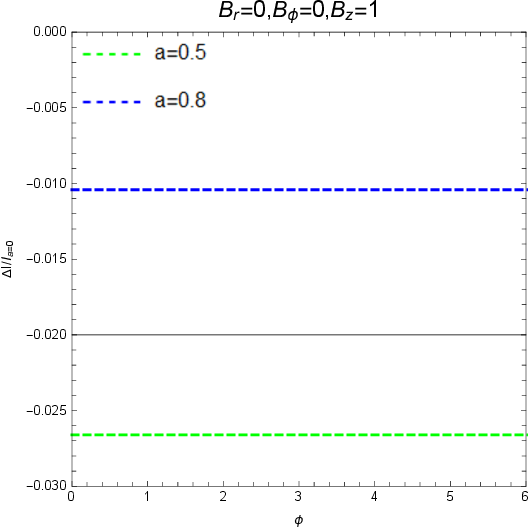}
\includegraphics[width=2.5cm,height=2.5cm]{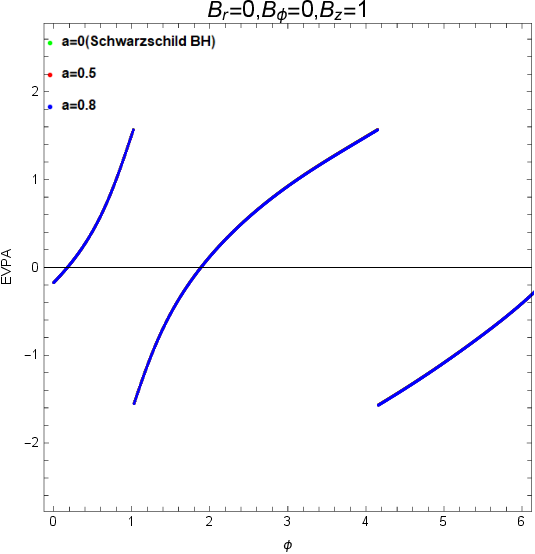}
\includegraphics[width=2.5cm,height=2.5cm]{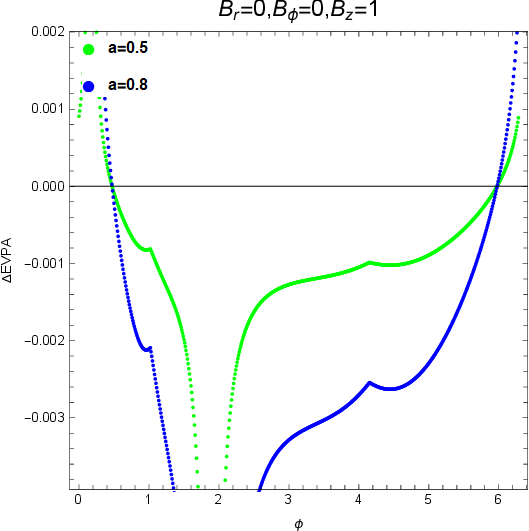}
\caption{The effects of magnetic fields on the polarized vector in the KS black hole spacetime for various deformation parameters $a$. {\em Top Panel}: Pure radial magnetic field. {\em Middle Panel}: Pure angular magnetic field. {\em bottom Panel}: Pure vertical magnetic field. Key parameters include a black hole mass of $M = 1$, an emitting ring radius of $R=6$, an observation angle $\theta_{0}=20^{\circ}$, a fluid velocity $\beta=0.3$, and a fluid direction angle $\chi=-90^{\circ}$.}
\label{fig:3}
\end{figure*}

\par
We also investigate the influence of magnetic fields oriented in different directions on polarization images. As depicted in Fig. \ref{fig:3}, when considering only the radial magnetic field, distinct maxima and minima in polarization intensity are observed, with the minimum occurring near $\phi=3.45\pi$ and the maximum near $\phi=0.35\pi$. The EVPA exhibits periodic variations with azimuthal changes. In the case of solely an azimuthal magnetic field, the maximum polarization intensity ($I_{\rm max}\simeq0.29$) increases, accompanied by a decrease in the corresponding phase angle. Conversely, the minimum polarization intensity ($I_{\rm min}\simeq0.105$) and its associated phase angle also decrease. Furthermore, an increase in deformation parameters corresponds to a decrease in the polarization intensity value. This implies that larger deformation parameters result in diminished spatiotemporal polarization. Through comparative analysis, we observed that the polarization outcomes of KS black holes resemble those of Schwarzschild black holes. We investigated whether exclusively a vertical magnetic field in spacetime produces a polarization pattern inconsistent with the observed results of M87$^{*}$. The observed increase in polarization intensity on the left side of the image due to aberration is attributed to the motion effect, a phenomenon inherent in any static spacetime.
\begin{figure*}[htbp]
\centering
\includegraphics[width=2.8cm,height=2.8cm]{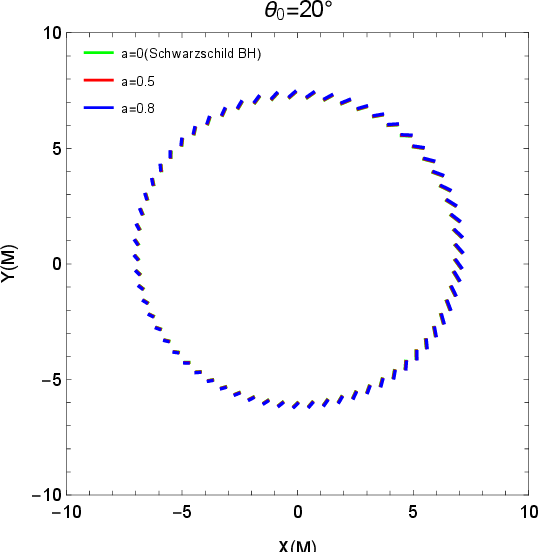}
\includegraphics[width=2.8cm,height=2.8cm]{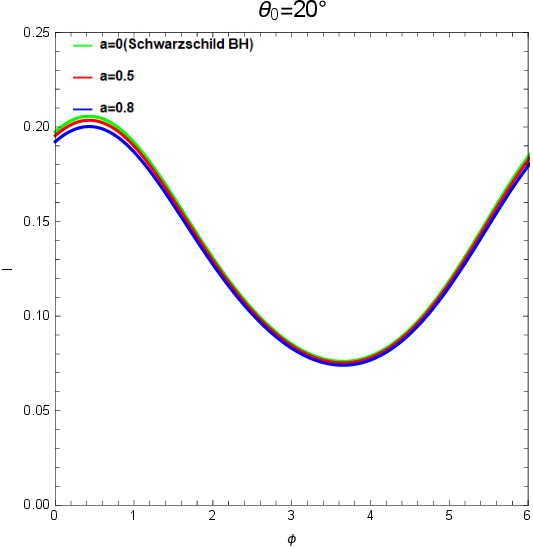}
\includegraphics[width=2.8cm,height=2.8cm]{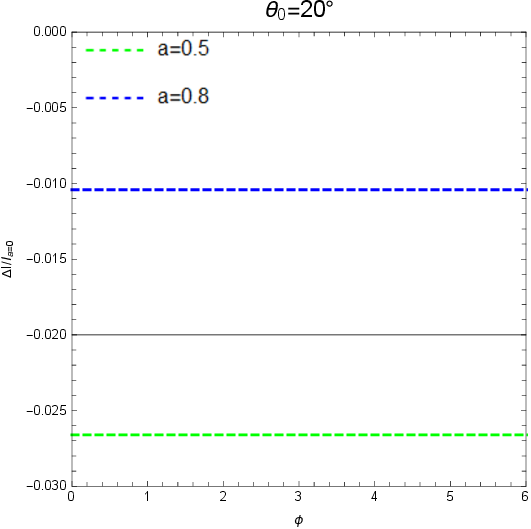}
\includegraphics[width=2.8cm,height=2.8cm]{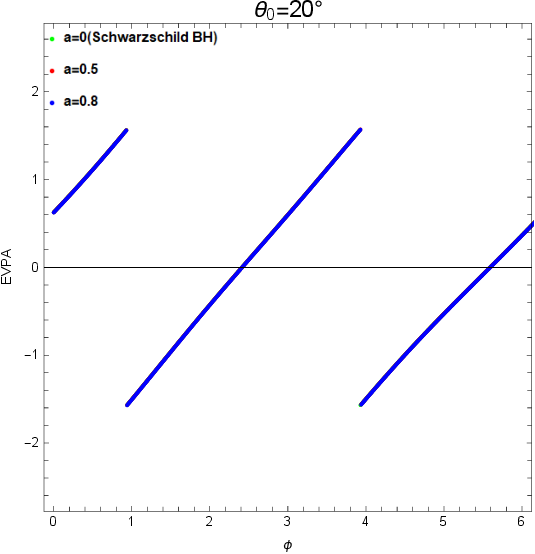}
\includegraphics[width=2.8cm,height=2.8cm]{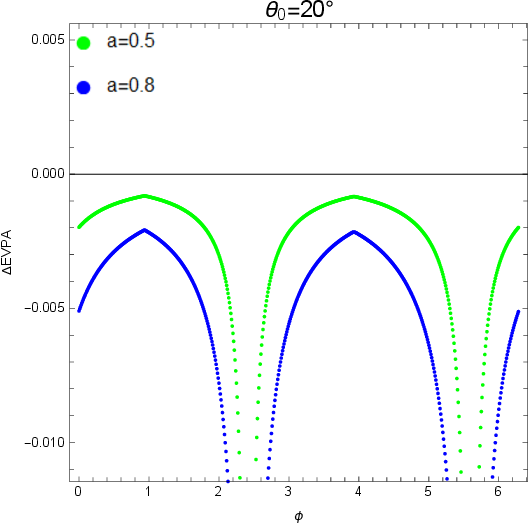}
\includegraphics[width=2.8cm,height=2.8cm]{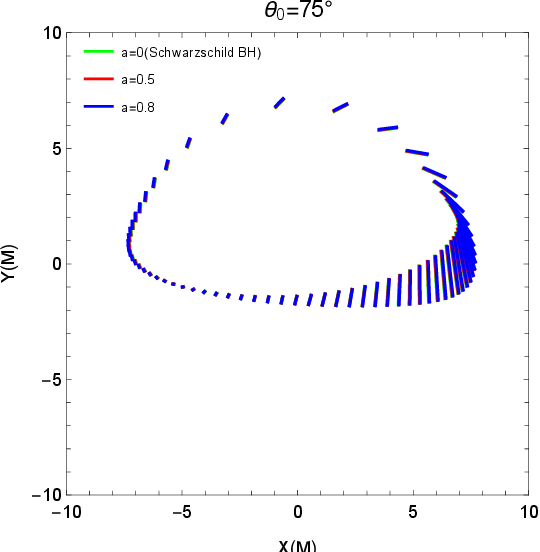}
\includegraphics[width=2.8cm,height=2.8cm]{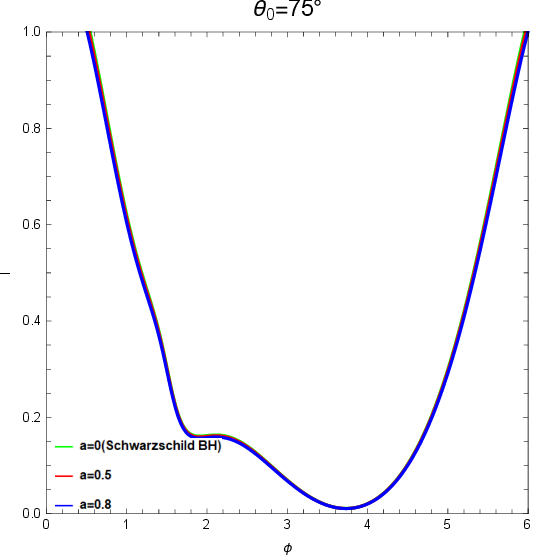}
\includegraphics[width=2.8cm,height=2.8cm]{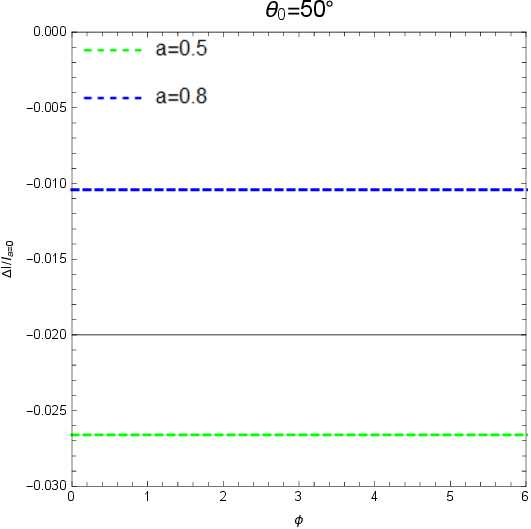}
\includegraphics[width=2.8cm,height=2.8cm]{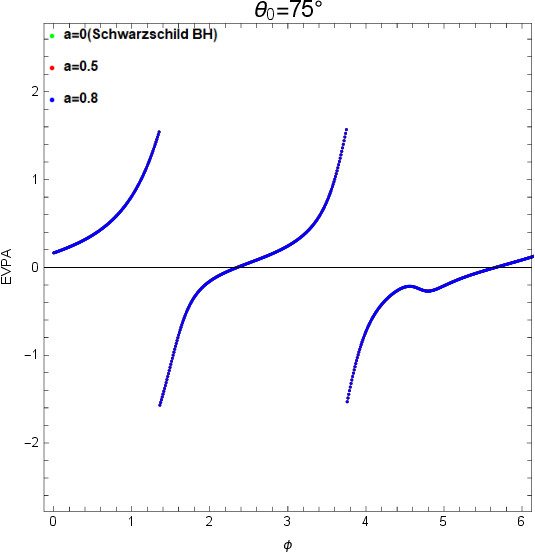}
\includegraphics[width=2.8cm,height=2.8cm]{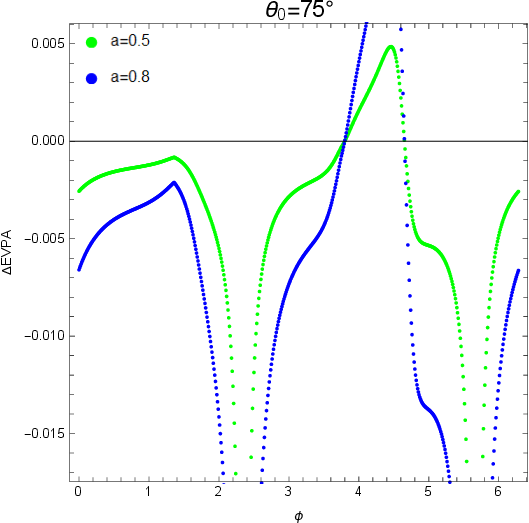}
\includegraphics[width=2.8cm,height=2.8cm]{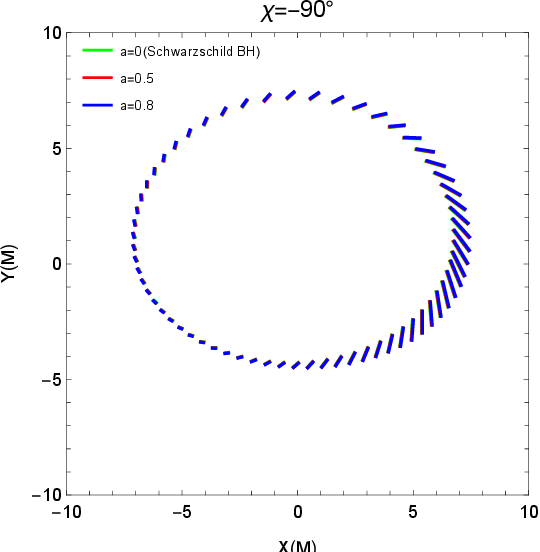}
\includegraphics[width=2.8cm,height=2.8cm]{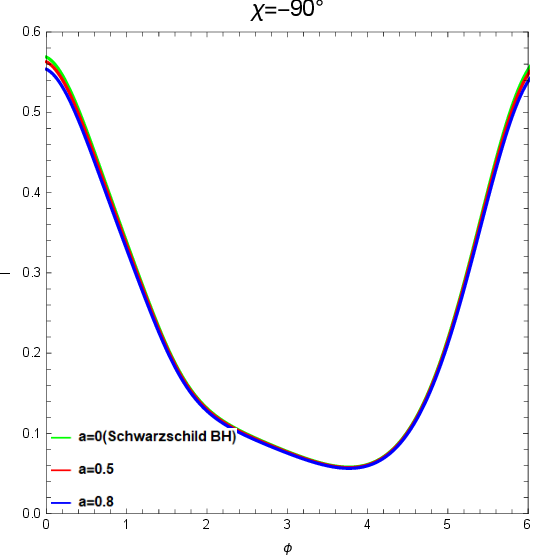}
\includegraphics[width=2.8cm,height=2.8cm]{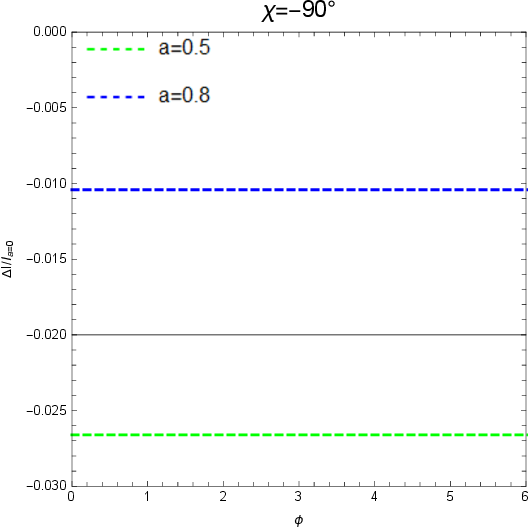}
\includegraphics[width=2.8cm,height=2.8cm]{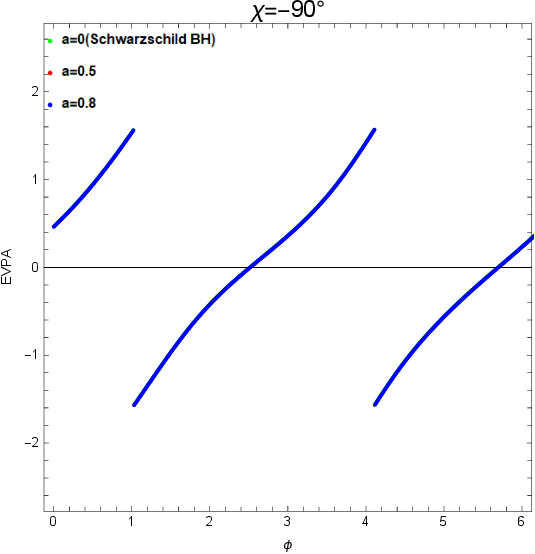}
\includegraphics[width=2.8cm,height=2.8cm]{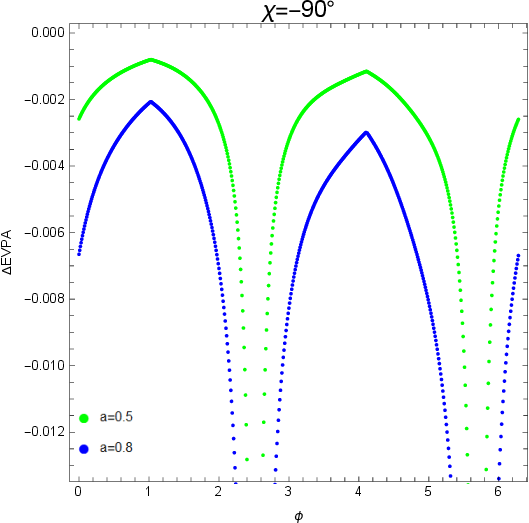}
\includegraphics[width=2.8cm,height=2.8cm]{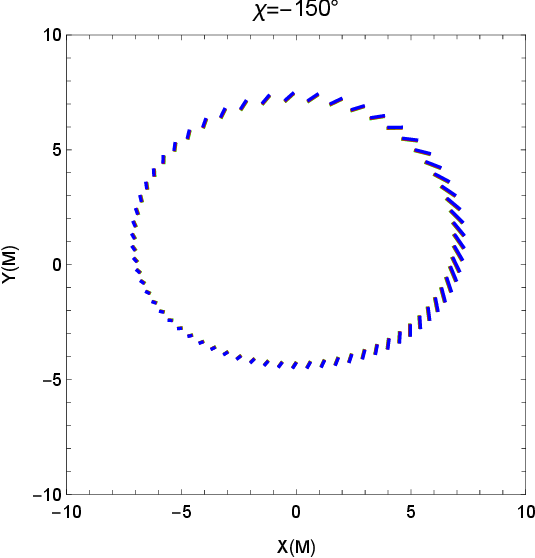}
\includegraphics[width=2.8cm,height=2.8cm]{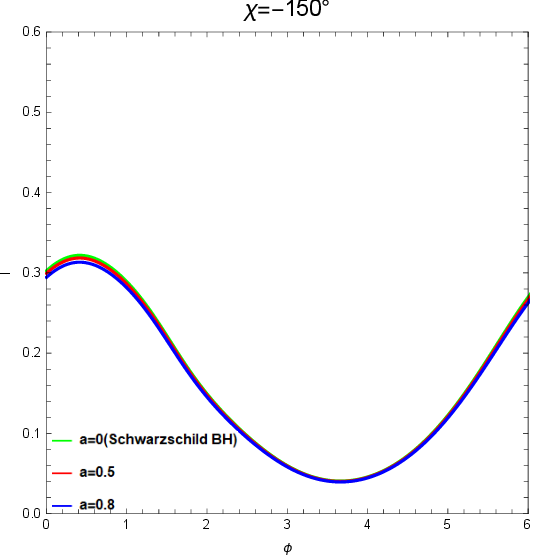}
\includegraphics[width=2.8cm,height=2.8cm]{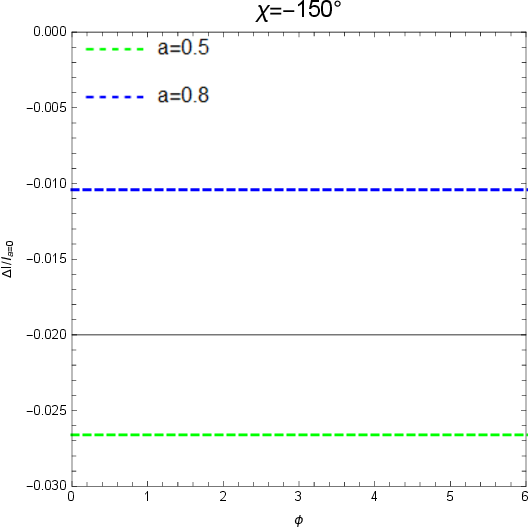}
\includegraphics[width=2.8cm,height=2.8cm]{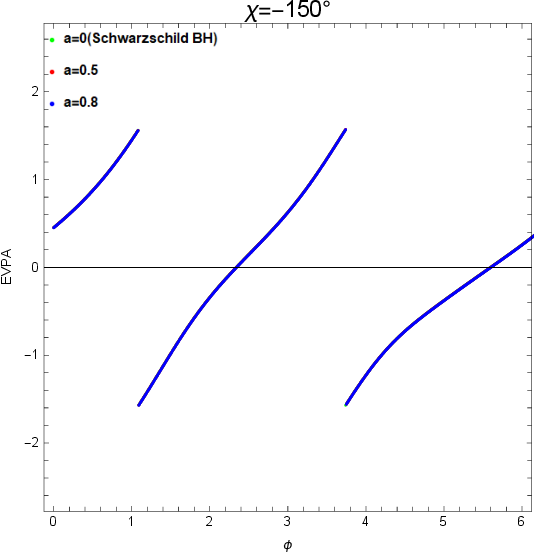}
\includegraphics[width=2.8cm,height=2.8cm]{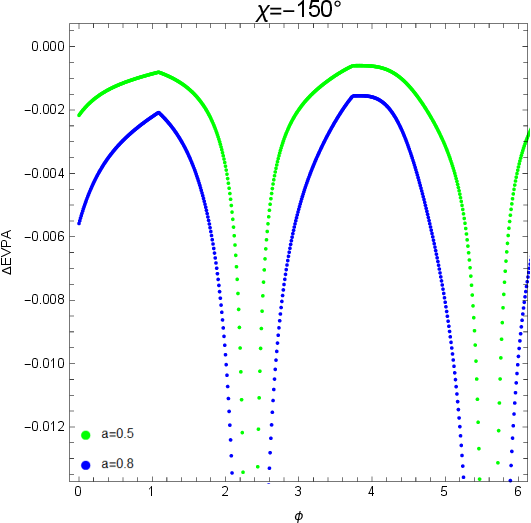}
\includegraphics[width=2.8cm,height=2.8cm]{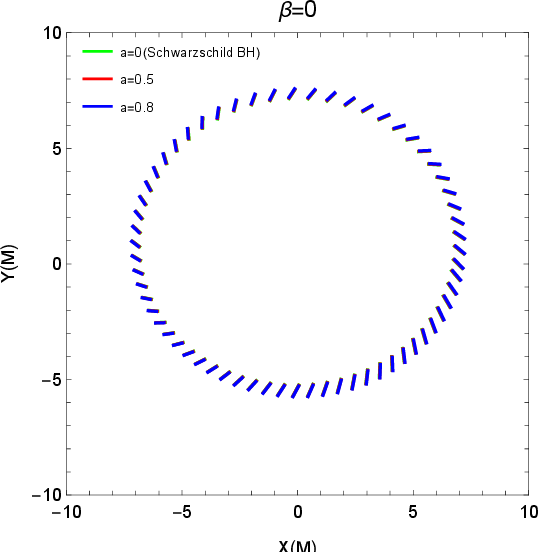}
\includegraphics[width=2.8cm,height=2.8cm]{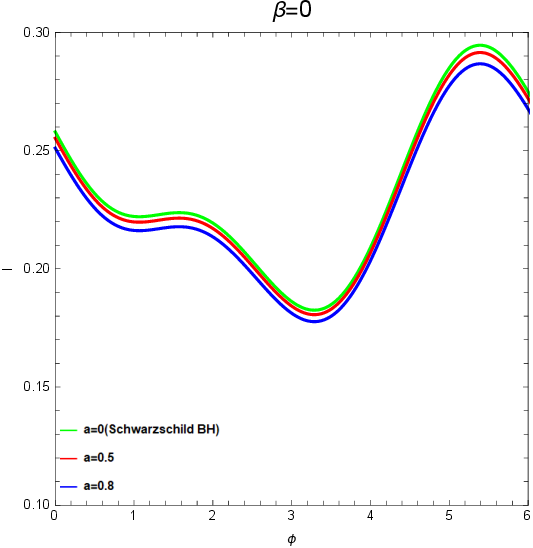}
\includegraphics[width=2.8cm,height=2.8cm]{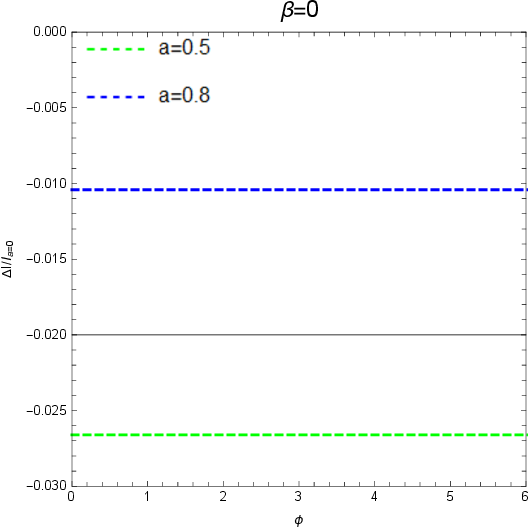}
\includegraphics[width=2.8cm,height=2.8cm]{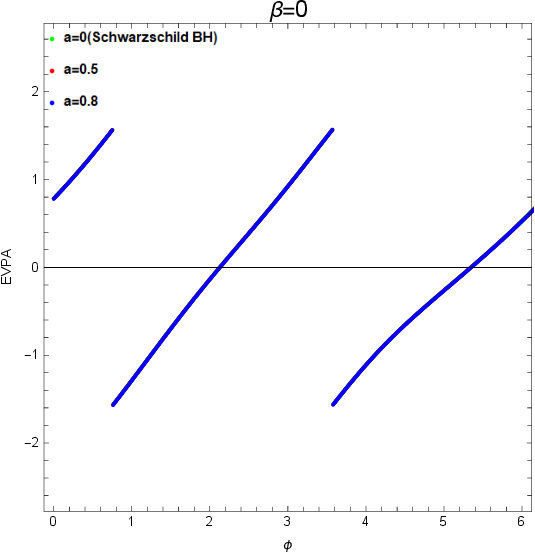}
\includegraphics[width=2.8cm,height=2.8cm]{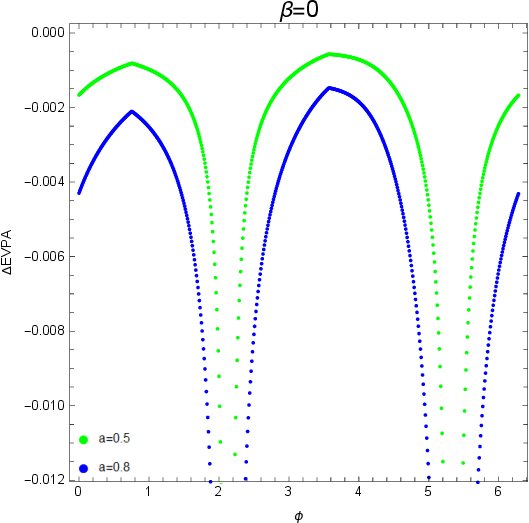}
\includegraphics[width=2.8cm,height=2.8cm]{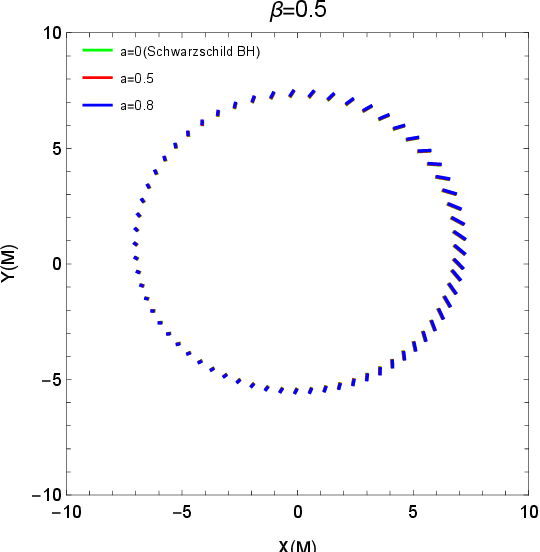}
\includegraphics[width=2.8cm,height=2.8cm]{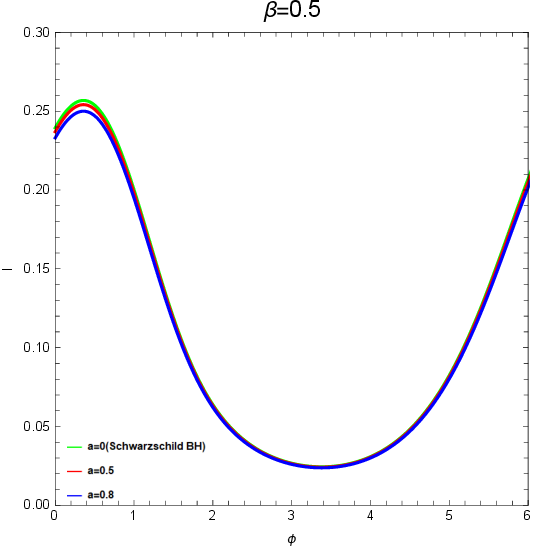}
\includegraphics[width=2.8cm,height=2.8cm]{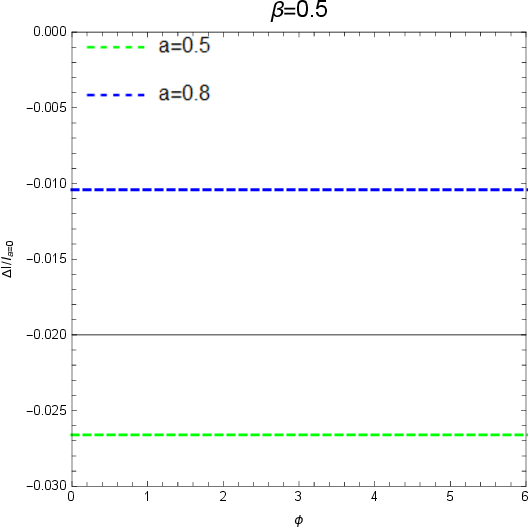}
\includegraphics[width=2.8cm,height=2.8cm]{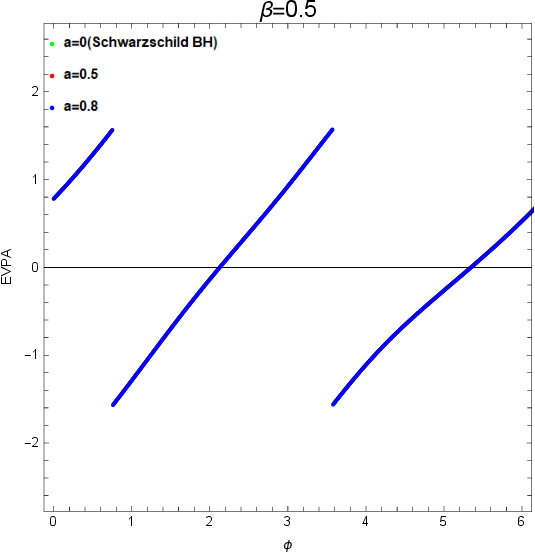}
\includegraphics[width=2.8cm,height=2.8cm]{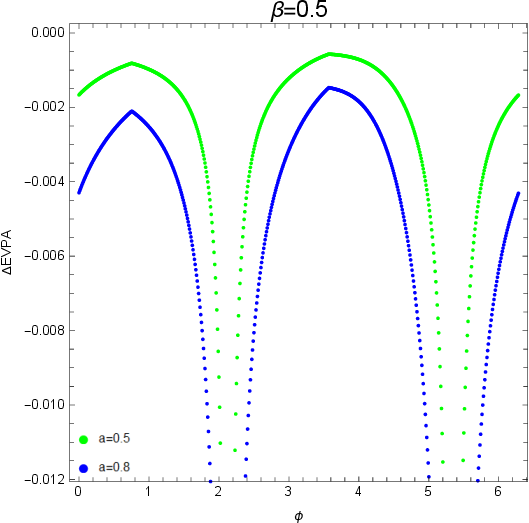}
\caption{The impact of observation angle, fluid direction angle, and fluid velocity on the polarized vector within the KS black hole under various deformation parameter values. {\em Top Two Panel}: Observation angles are set to $\theta_{0}=20^{\circ}$ and $\theta_{0}=75^{\circ}$. The parameters for these panels include a black hole mass of $M=1$, a fluid direction angle $\chi=-135^{\circ}$, an emitting ring radius of $R=6$, a fluid velocity of $\beta=0.3$, and a magnetic field configuration of $B_{r}=0.87$, $B_{\phi}=0.5$, and $B_{z}=0$. {\em Middle Two Panel}: Fluid direction angles are specified as $\chi=-90^{\circ}$ and $\chi=-150^{\circ}$. The associated parameters are $M=1$, $\theta_{0}=45^{\circ}$, $R=6$, $\beta=0.3$, $B_{r}=0.71$, $B_{\phi}=0.71$, and $B_{z}=0$. {\em Bottom Two Panel}: Fluid direction angles are presented as $\beta=0$ and $\beta=0.5$. The parameters for these panels include $M=1$, $\theta_{0}=30^{\circ}$, $R=6$, $\chi=-120^{\circ}$, $B_{r}=0.5$, $B_{\phi}=0.87$, and $B_{z}=0$.}
\label{fig:4}
\end{figure*}

\par
We present a polarization image capturing the angular and radial components of the equatorial magnetic field. In Fig. \ref{fig:4}, the notable observation is the substantial increase in polarization intensity with rising observation angles. Particularly, at $\theta_{0}=75^{\circ}$, the maximum polarization intensity ($I_{\rm max}\simeq1$) is approximately five times that observed at $\theta_{0}=20^{\circ}$ ($I_{\rm max}\simeq0.2$). This observation implies that enhancing observation inclination angles in future studies can augment the accuracy of polarization data. Conversely, polarization intensity experiences a decline with increasing deformation parameter. As the observation tilt angle rises, the linear correlation between EVPA changes and the azimuthal period $\phi$ diminishes, resulting in non-linear distortions. This intricate relationship suggests that heightened tilt angles induce irregular distortions in EVPA, deviating from the patterns observed in standard Schwarzschild black holes. This outcome serves as a potential indicator to discriminate between Schwarzschild black holes and quantum corrected black hole. It is noteworthy that our analysis is conducted under the condition of $B_{\rm r} > B_{\rm \phi}$.

\par
In the scenario where $B_{\rm r} = B_{\rm \phi}$, an examination reveals that when $\chi$ equals $-90^{\circ}$, the polarization intensity curve mirrors that of a pure angular magnetic field. Conversely, when $\chi$ decreases to $-150^{\circ}$, the polarization intensity curve adopts a resemblance to that of a pure radial magnetic field. Notably, as the fluid direction angles decrease, the peak polarization intensity experiences a reduction, and the alteration in EVPA remains relatively modest. This phenomenon is attributed to the primary distortion of EVPA being predominantly influenced by magnetic fields, with the contribution from fluid direction angles being comparatively minor.

\par
We also investigate the scenario where $B_{\phi}$ exceeds $B_{r}$. Under zero-flow conditions, the polarization intensity exhibits an irregular curve, accompanied by quasi-periodic changes in the EVPA. Upon increasing the flow rate to $0.5$, the polarization intensity curve takes on a form resembling that of an exclusively radial magnetic field, accompanied by a concurrent decrease in the peak polarization intensity. Remarkably, the EVPA remains unchanged throughout this transition. Of particular interest is the constancy of $\Delta I$ \Big($\frac{I_{\rm a=0.5/0.8}-I_{\rm a=0}}{I_{\rm a=0}}$\Big) across all considered cases. This suggests that regardless of alterations in external conditions, the discrepancy in polarization intensity remains unaltered, with larger deformation parameters approaching $0$. Simultaneously, an increase in deformation parameters corresponds to a reduction in polarization intensity, unveiling an anti-polarization effect not previously observed in prior studies.

\par
\begin{figure*}[ht]
\centering
\includegraphics[width=4.5cm,height=4.5cm]{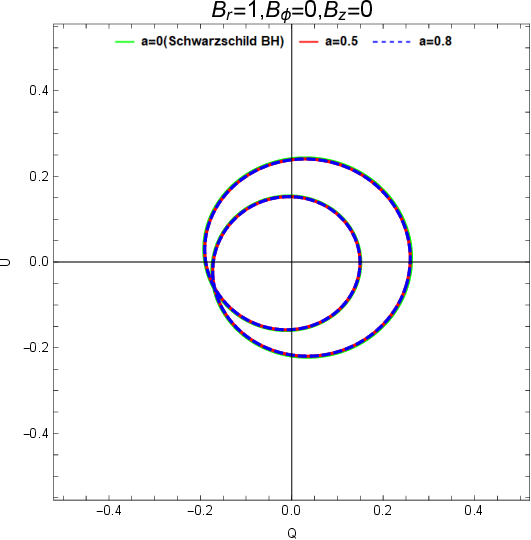}
\hspace{0.1cm}
\includegraphics[width=4.5cm,height=4.5cm]{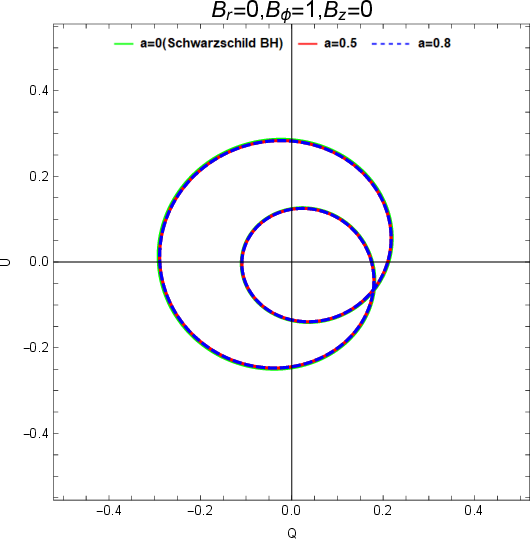}
\hspace{0.1cm}
\includegraphics[width=4.5cm,height=4.5cm]{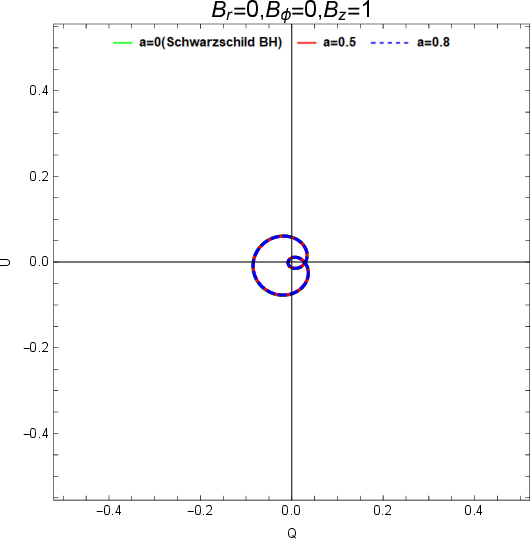}
\hspace{0.1cm}
\includegraphics[width=4.5cm,height=4.5cm]{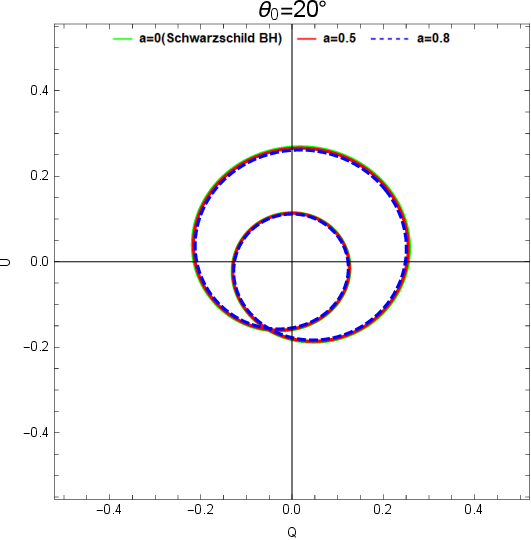}
\hspace{0.1cm}
\includegraphics[width=4.5cm,height=4.5cm]{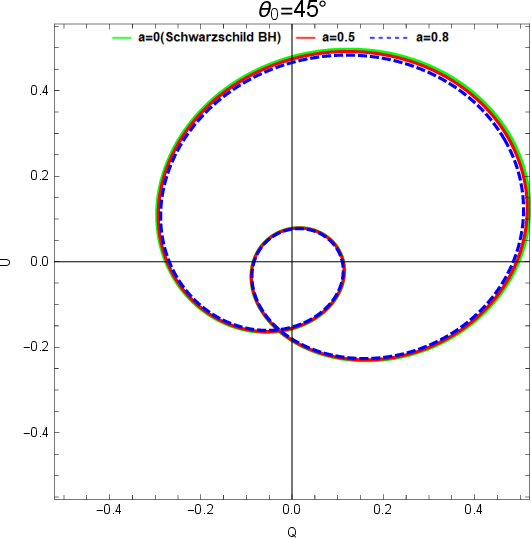}
\hspace{0.1cm}
\includegraphics[width=4.5cm,height=4.5cm]{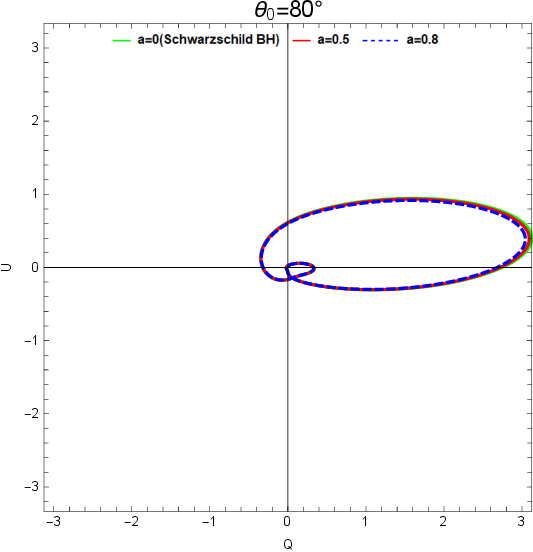}
\hspace{0.1cm}
\includegraphics[width=4.5cm,height=4.5cm]{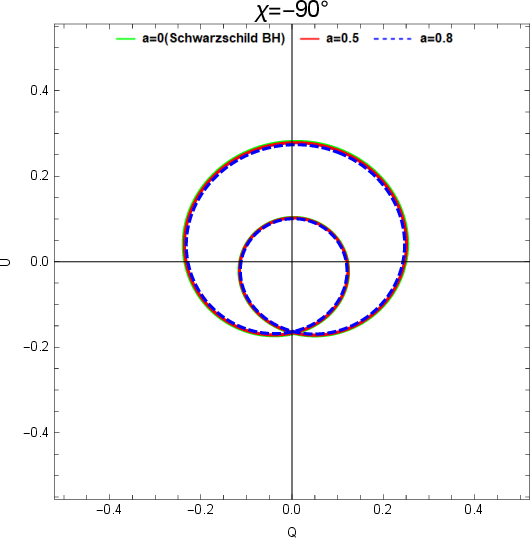}
\hspace{0.1cm}
\includegraphics[width=4.5cm,height=4.5cm]{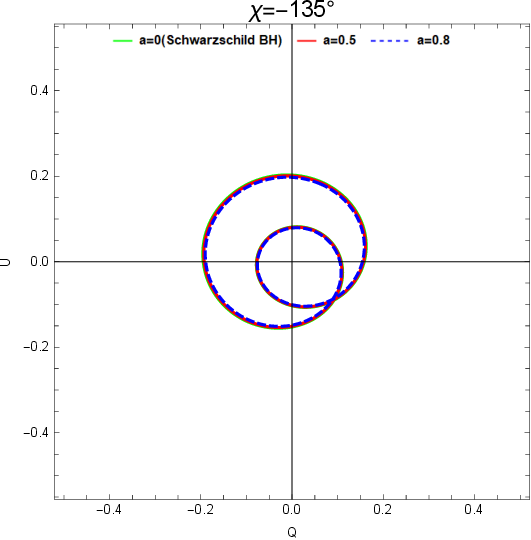}
\hspace{0.1cm}
\includegraphics[width=4.5cm,height=4.5cm]{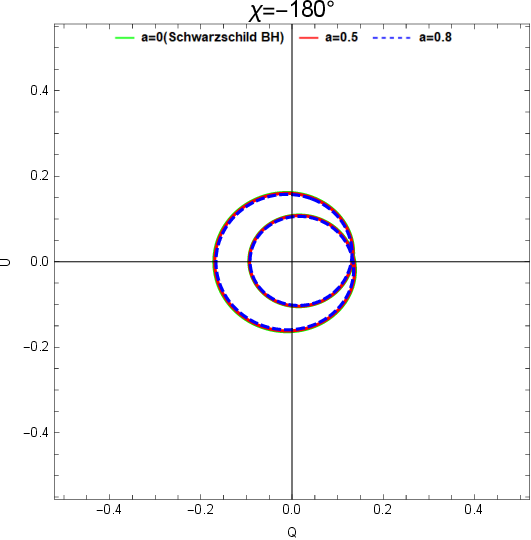}
\caption{Impact of varied deformation parameters in the KS black hole on the $U-Q$ plane. {\em Top Panel}: different magnetic fields. The parameters are set to $M=1$, $\chi=-90^{\circ}$, $R=6$, $\beta=0.3$, and $\theta_{0}=20^{\circ}$. {\em Middle Panel}: different observation angles. The parameters for this scenario include $M=1$, $\chi=-90^{\circ}$, $R=6$, $\beta=0.3$, $B_{\rm r}=0.5$, $B_{\rm \phi}=0.87$, and $B_{\rm z}=0$. {\em Bottom Panel}: different fluid direction angles. Parameters for this investigation encompass $M=1$, $R=6$, $\beta=0.3$, $B_{\rm r}=0.71$, $B_{\rm \phi}=0.71$, $B_{\rm z}=0$, and $\theta_{0}=20^{\circ}$.}
\label{fig:5}
\end{figure*}

Finally, our investigation focused on the $U-Q$ loop diagram representing the polarization vectors in the emission ring image encircling a KS black hole, elucidating the continuous variability of polarization vectors in the emission ring image around the black hole. As depicted in Fig. \ref{fig:5}, akin to conventional static black holes, two rings manifest around the origin of the black hole in the $U-Q$ plane. Our examination encompassed scenarios involving magnetic fields in the radial, angular, and vertical directions. In cases of pure radial and angular magnetic fields, a reversal in the direction of the $U-Q$ ring plot was observed. Conversely, when exclusively considering the vertical magnetic field, the two loops markedly contracted towards the center. With an increase in the observer's inclination angle, the inner ring exhibited gradual reduction, ultimately contracting to a point and then vanishing completely. As the observer's inclination angle increased from $\theta_{0}=0^{\circ}$ to $\theta_{0}=180^{\circ}$, the orientation of the magnetic rings altered, and both rings contracted inwards. Upon scrutiny of the $U-Q$ diagrams under diverse conditions, the impact of deformation parameters was found to be relatively minor. However, with an escalation in the deformation parameter, the $U-Q$ ring exhibited an inward contraction.

\section{Polarized image of KS black hole and compared with M87$^{*}$}
\label{sec:4}
In this section, we scrutinize the KS black hole surrounded by a thin accretion disk in the equatorial plane, deriving the corresponding observable image. We proceed under the assumption that the fluid particles within the disk adhere to geodesic circular orbits. The electromagnetic radiation flux, denoted in units of $\rm ergs^{-1}cm^{-2}str^{-1}Hz^{-1}$, emanates from a specified radial position $r$ on the accretion disk. According to ref. \cite{42}, we ascertain
\begin{equation}
\label{4-1}
F = - \frac{\dot{M}}{4\pi \sqrt{\rm -I}} \frac{\Omega_{,\rm r}}{(E-\Omega L)^{2}} \int_{r_{\rm in}}^{r} (E- \Omega L)L_{,\rm r} {\rm d} r,
\end{equation}
where $\dot{M}$ denotes the mass accretion rate, $I$ denotes the determinant of the induced metric in the equatorial plane, and $r_{\rm in}$ denotes the inner edge of the accretion disk. The quantities $E$, $\Omega$, and $L$ represent, respectively, the angular velocity, energy, and angular momentum of particles in a circular orbit. These parameters can be expressed as follows:
\begin{align}
\label{4-2}
&E= - \frac{g_{\rm tt} + g_{\rm t \phi} \Omega}{\sqrt{-g_{\rm tt} + 2g_{\rm t \phi}\Omega - g_{\rm \phi \phi}\Omega^{2}}},\\
\label{4-3}
&L= \frac{g_{\rm t \phi} + g_{\rm \phi \phi} \Omega}{\sqrt{-g_{\rm tt} + 2g_{\rm t \phi}\Omega - g_{\rm \phi \phi}\Omega^{2}}},\\
\label{4-4}
&\Omega=\frac{{\rm d}\phi}{{\rm d}t}=\frac{-g'_{\rm t \phi} + \sqrt{(g'_{\rm t \phi})^{2} - g'_{\rm tt}g'_{\rm \phi \phi}}}{g'_{\rm \phi \phi}}.
\end{align}

\par
\begin{figure*}[ht]
\centering
\includegraphics[width=4.5cm,height=3.7cm]{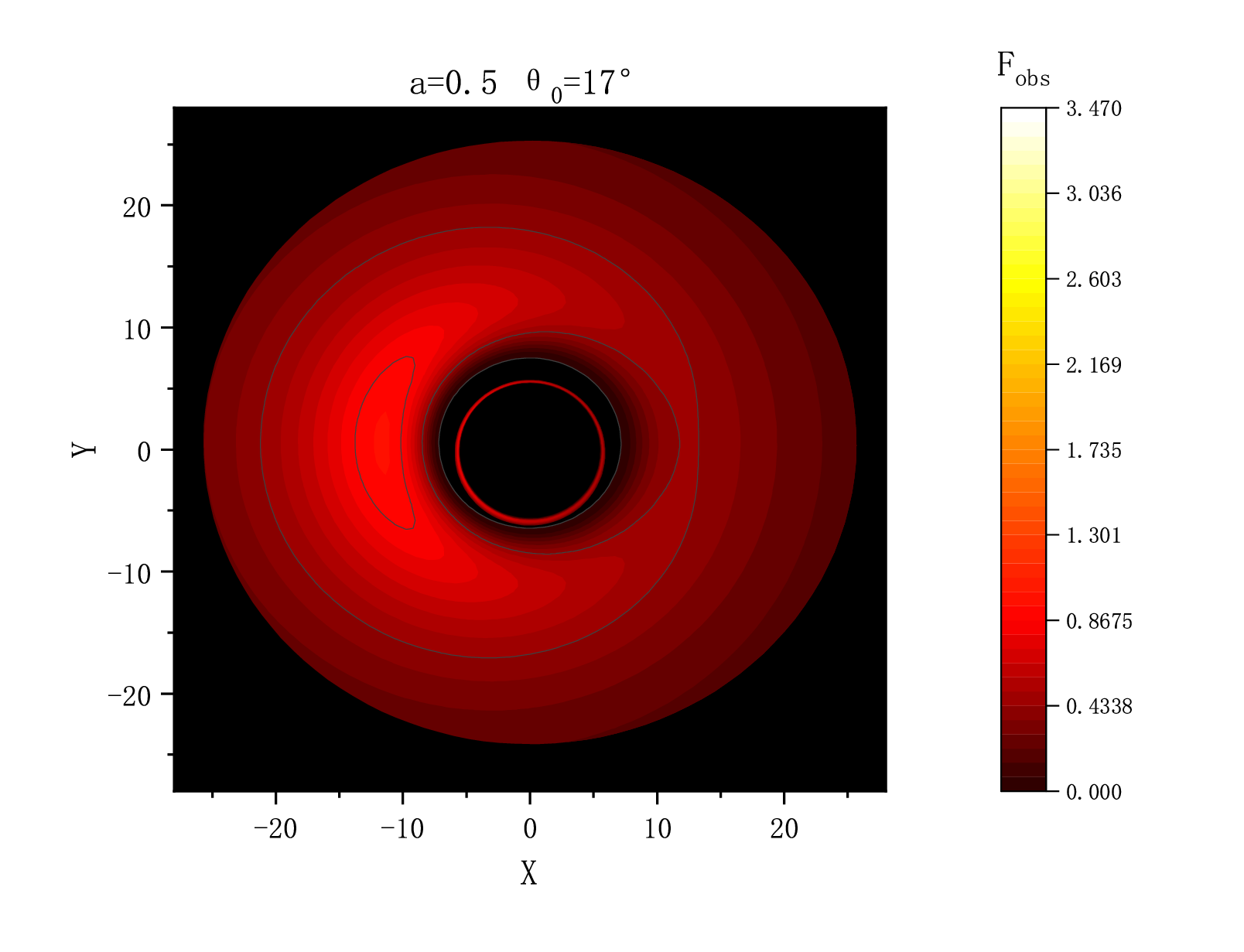}
\hspace{0.1cm}
\includegraphics[width=4.5cm,height=3.7cm]{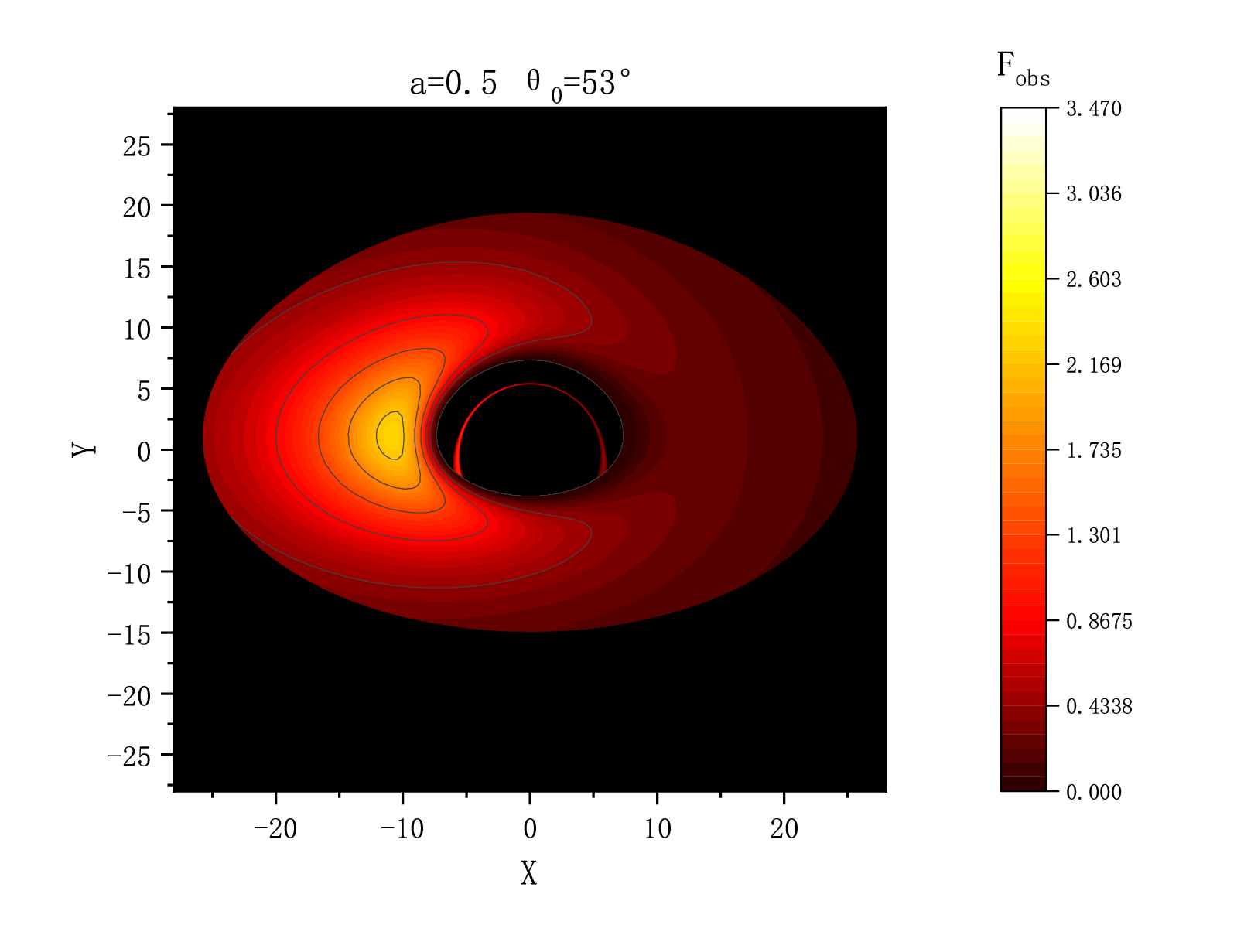}
\hspace{0.1cm}
\includegraphics[width=4.5cm,height=3.7cm]{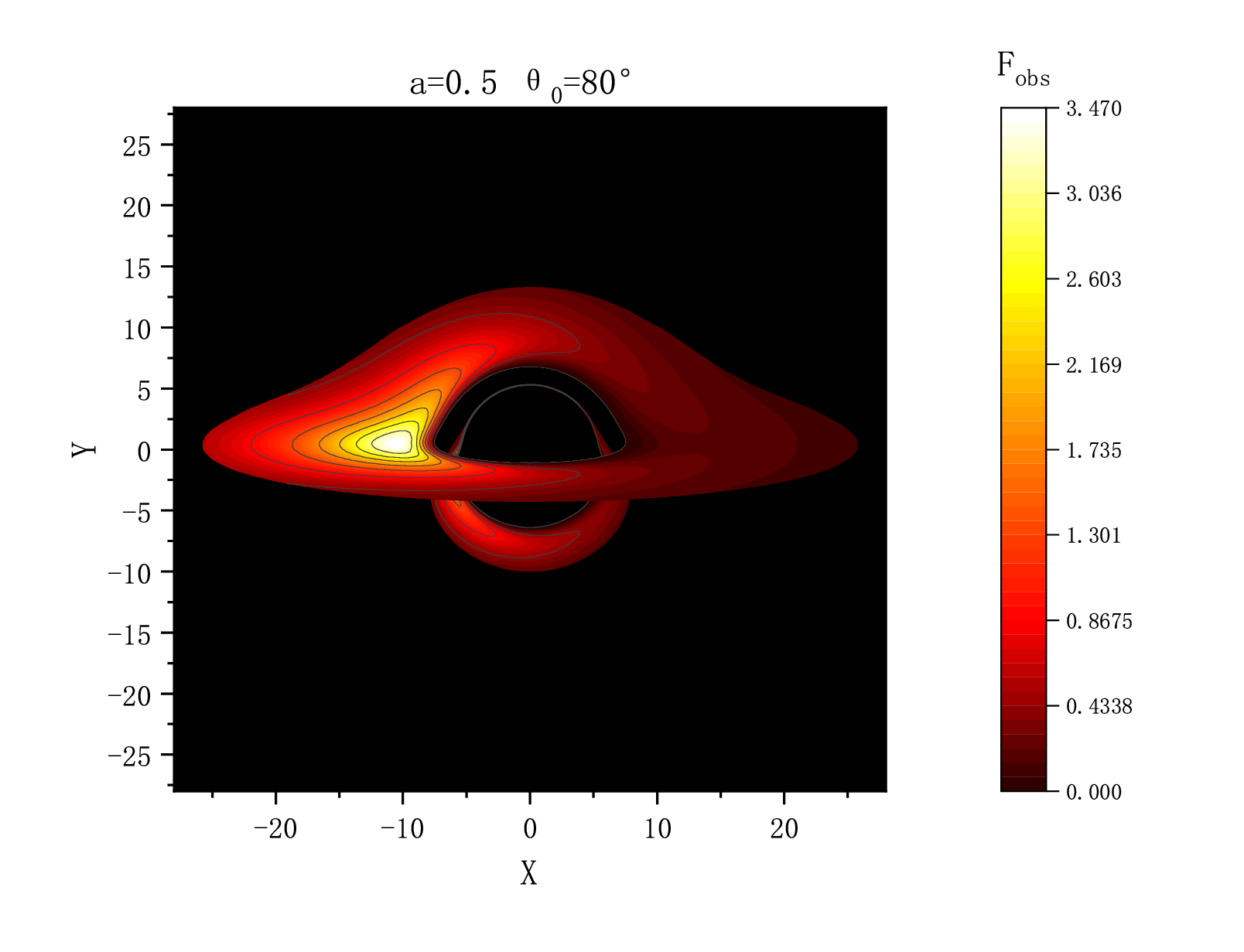}
\hspace{0.1cm}
\includegraphics[width=4.5cm,height=3.7cm]{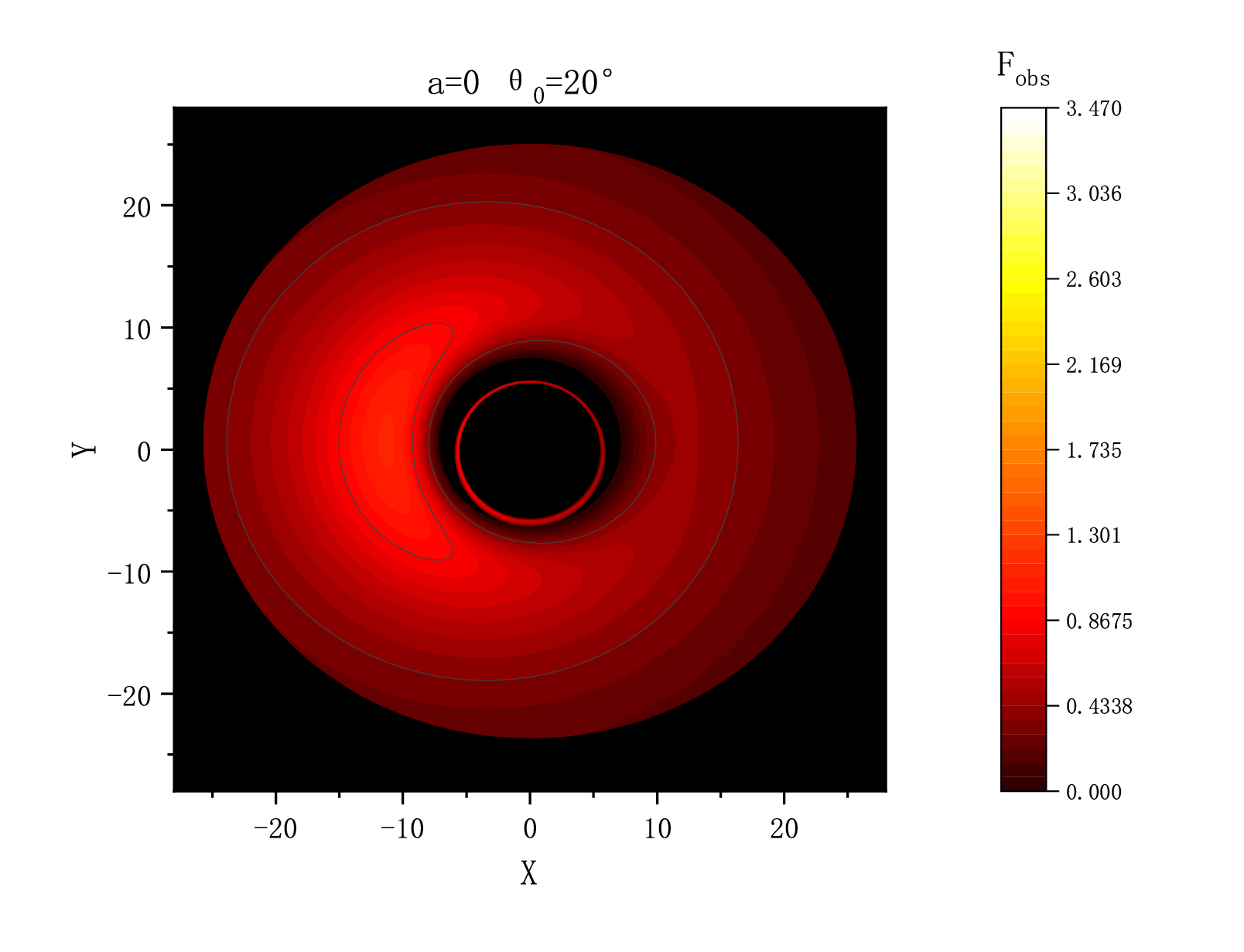}
\hspace{0.1cm}
\includegraphics[width=4.5cm,height=3.7cm]{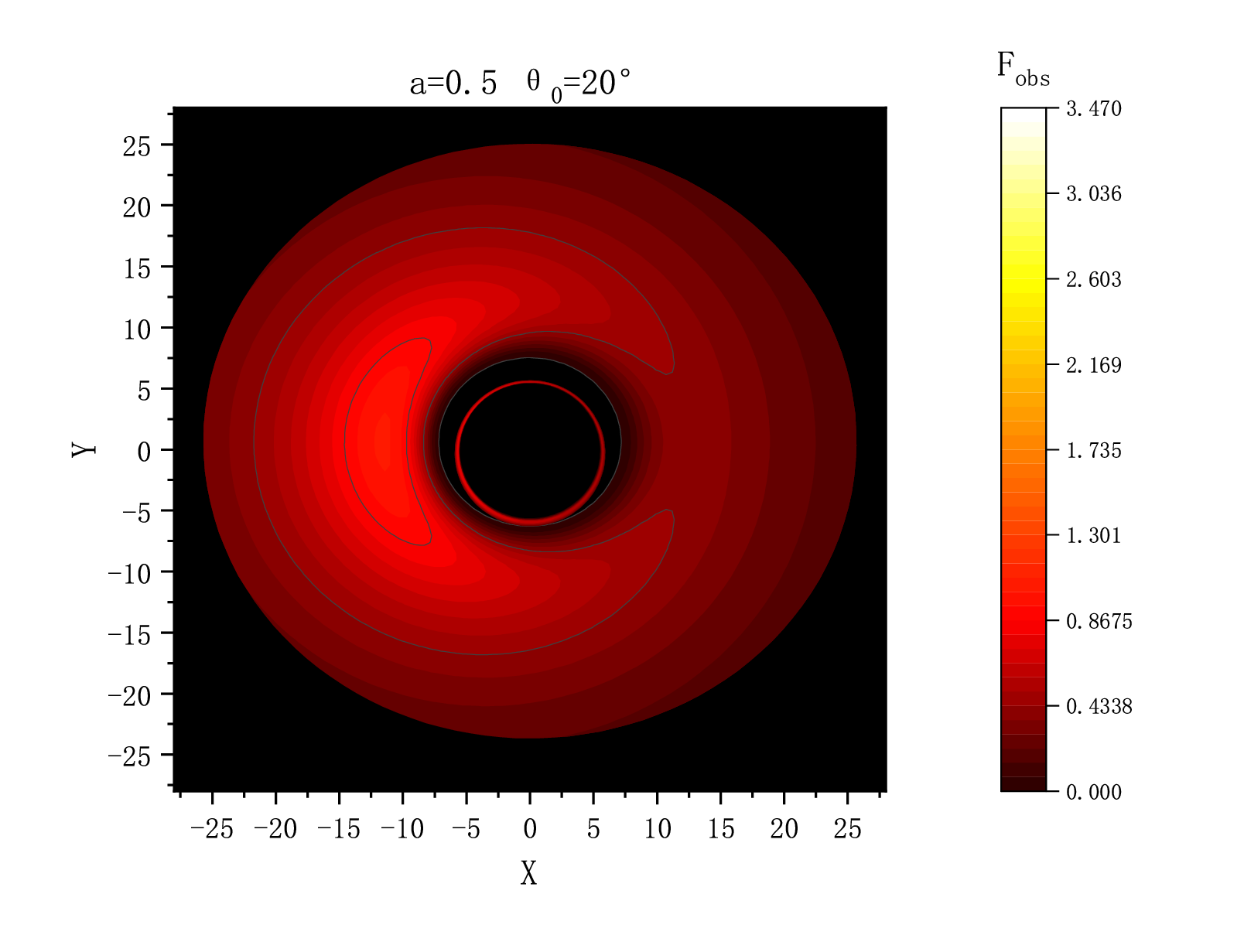}
\hspace{0.1cm}
\includegraphics[width=4.5cm,height=3.7cm]{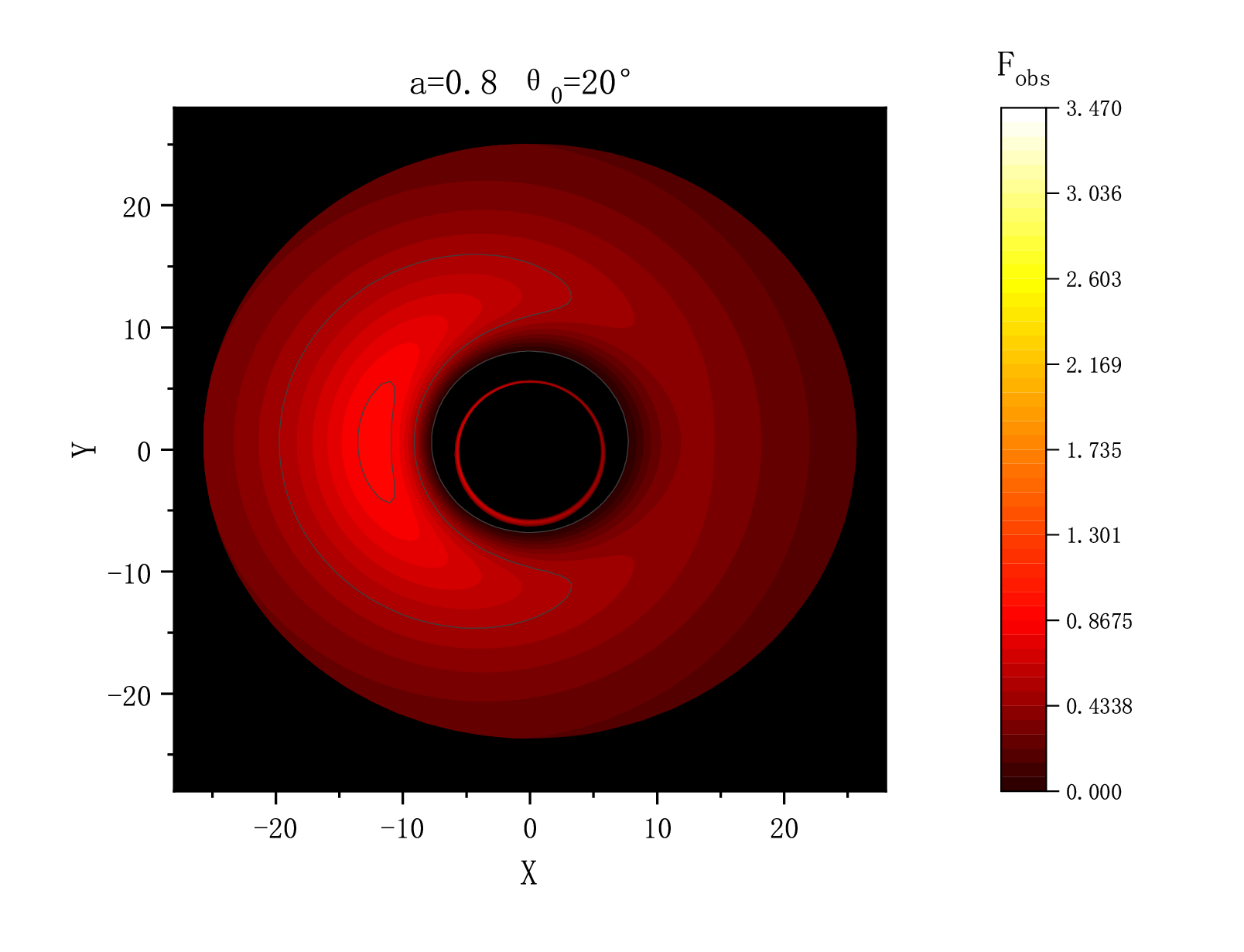}
\caption{Observation images of a KS black hole within accretion disks at varying observation angles. {\em Top Left Panel}: Observation angle $\theta_{0}=17^{\circ}$. {\em Top Middle Panel}: Observation angle $\theta_{0}=53^{\circ}$. {\em Top Right Panel}: Observation angle $\theta_{0}=80^{\circ}$. {\em Bottom Panel}: Observation angle $\theta_{0}=20^{\circ}$ and deformation parameter $a=0,0.5,0.8$.}
\label{fig:6}
\end{figure*}

The observational images of a KS black hole within accretion disks context at different observation angles are depicted in Fig. \ref{fig:6}. An increase in the observation inclination angle results in the configuration assuming a hat-like shape, characterized by a noticeable concentration of luminosity, particularly on the left side of the accretion disk. Variations in the deformation parameters reveal that an increase in the parameter $a$ correlates with a significant reduction in the accumulation of luminosity on the left side (quantified by the area of contour lines within the brightness region), attributable to quantum correction effects \cite{38}.

\par
We aimed to validate the ability of KS black holes to reproduce polarization images resembling those observed in M87$^{*}$. To achieve this, we carefully selected physical parameters consistent with observed data to capture the primary qualitative characteristics of the radio source. The M87$^{*}$ image exhibits a distorted polarization pattern, featuring a $72^{\circ}$ rotation of the polarization vector around the ring. Additionally, the strong flux is concentrated on the right side of the image, gradually diminishing towards the north and south poles. Previous investigations focusing on polarization ring models within the context of Schwarzschild black holes have provided insights into the impact of various parameters. The polarization mode is primarily influenced by the direction of the magnetic field, imposing stringent constraints on its orientation. For Schwarzschild black hole, the most suitable magnetic field direction is estimated to be $B_{(\rm r,\phi,z)}=(0.87,0.5,0)$. Consequently, we chose to simulate the same magnetic field configuration and conducted a comparative analysis.

\par
\begin{figure*}[ht]
\centering
\includegraphics[width=3.5cm,height=3.2cm]{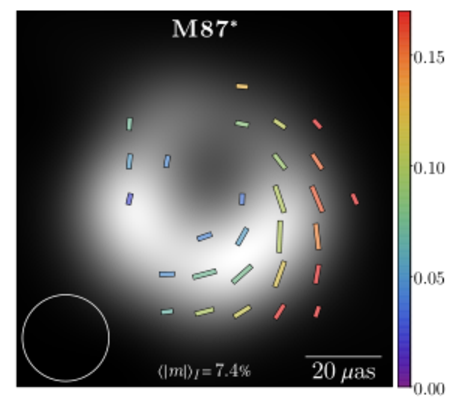}
\hspace{0.1cm}
\includegraphics[width=3cm,height=3.2cm]{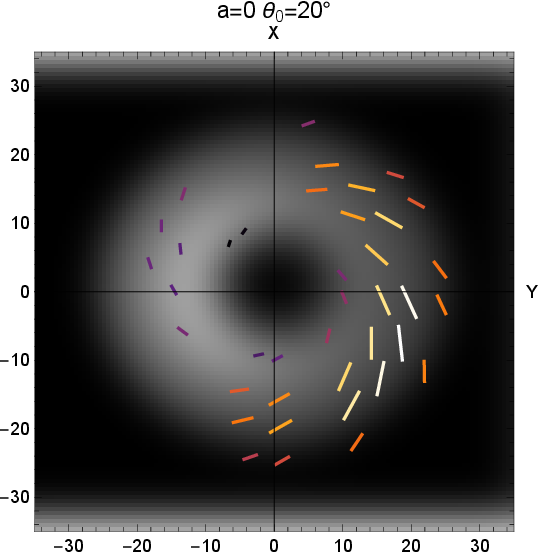}
\hspace{0.1cm}
\includegraphics[width=3cm,height=3.2cm]{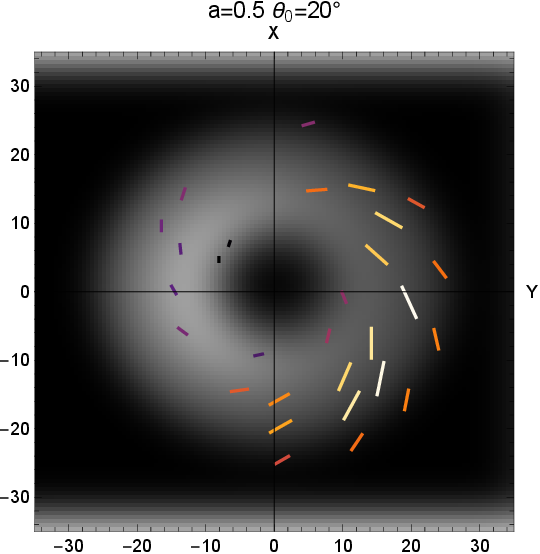}
\hspace{0.1cm}
\includegraphics[width=3cm,height=3.2cm]{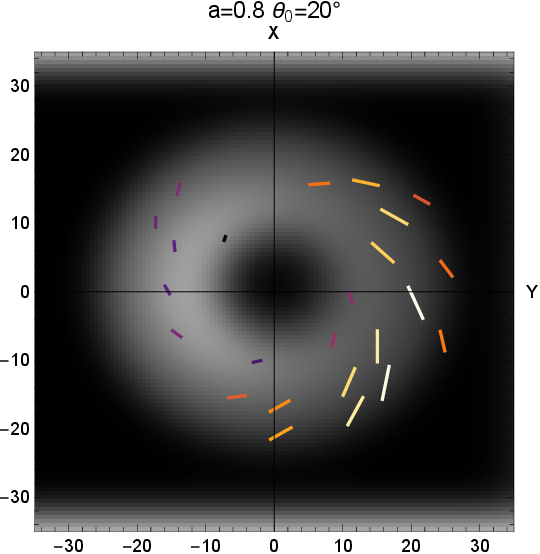}
\caption{Comparison between the polarimetric image of KS black hole and the M87$^{*}$ black hole. Here $M=1$, $R=6$, $\beta=0.4$, $\chi=-150^{\circ}$, $B_{r}=0.87$, $B_{\phi}=0.5$, $B_{z}=0$, and $\theta_{0}=17^{\circ}$}
\label{fig:7}
\end{figure*}

In Fig. \ref{fig:7}, we juxtaposed the polarization images of KS black hole with those of M87$^{*}$. The observations reveal that the polarization images of KS black hole, with distinct deformation parameters, exhibit spiral structures reminiscent of M87$^{*}$. Moreover, the polarization flux of KS black holes is similarly concentrated on the right side. However, notable differences were observed, particularly with the increase in deformation parameters leading to a gradual expansion of the polarization region. In comparison to the M87$^{*}$ black hole, the flux density has increased, suggesting that M87$^{*}$ may not belong to the KS black hole category. This underscores that the depiction of a black hole's image depends not only on the black hole's inherent properties but also on the surrounding material and the corresponding radiation.
\begin{table*}[htbp]
\footnotesize
\caption{Numerical relationship between polarization intensity and deformation parameter for $B_{(\rm r,\phi,z)}=0.87,0.5,0$.}
\label{Tab:1}
\tabcolsep 18pt
\begin{tabular}{c|cccccccccccccccccccccc}
  \hline
  $a$   &   ${R=4.5M}$   &   ${R=6M}$  \\
  \hline
  $0.05$  &  $0.061761$   &$0.113913$  \\
  \hline
  $0.10$  &  $0.061721$   &$0.113877$  \\
 \hline
  $0.15$  &  $0.061651$   &$0.113818$  \\
  \hline
  $0.20$  &  $0.061555$   &$0.113735$  \\
  \hline
  $0.25$  &  $0.061432$   &$0.113628$  \\
  \hline
  $0.30$  &  $0.061281$   &$0.113497$  \\
  \hline
  $0.35$  &  $0.061103$   &$0.113343$  \\
  \hline
  $0.40$  &  $0.060898$   &$0.113165$  \\
  \hline
  $0.45$  &  $0.060665$   &$0.112964$  \\
  \hline
  $0.50$  &  $0.060405$   &$0.112739$  \\
  \hline
  $0.55$  &  $0.060119$   &$0.112492$  \\
  \hline
  $0.60$  &  $0.059805$   &$0.112218$  \\
  \hline
  $0.65$  &  $0.059464$   &$0.111922$  \\
  \hline
  $0.70$  &  $0.059097$   &$0.111602$  \\
  \hline
  $0.75$  &  $0.058703$   &$0.111261$  \\
  \hline
  $0.80$  &  $0.058283$   &$0.110893$  \\
  \hline
  $0.85$  &  $0.057836$   &$0.110503$  \\
  \hline
  $0.90$  &  $0.057363$   &$0.110091$  \\
  \hline
  $0.95$  &  $0.056864$   &$0.109654$  \\
  \hline
\end{tabular}
\end{table*}
\begin{figure}[h]
\centering
\includegraphics[width=9cm,height=5cm]{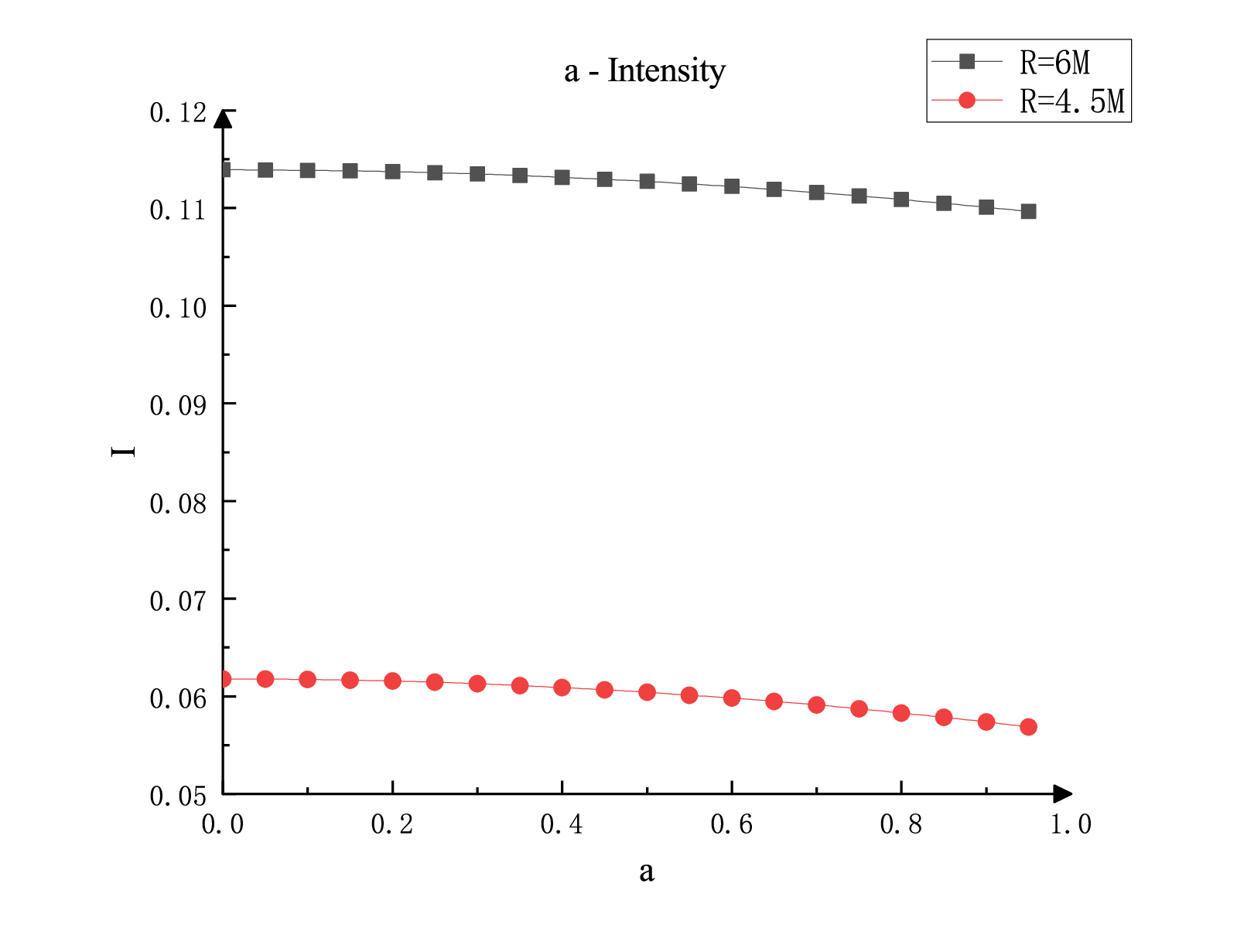}
\caption{The influence of deformation parameters on polarization intensity. Here $M = 1$, $\beta=0.4$, $\chi=-150^{\circ}$, $B_{r}=0.87$, $B_{\phi}=0.5$, $B_{z}=0$, and $\theta_{0}=17^{\circ}$}
\label{fig:8}
\end{figure}

\par
To highlight the influence of deformation parameters on polarization images, we compute the polarization intensity magnitude across various radii and deformation parameters (see Table \ref{Tab:1}). Subsequently, we establish the relationship between polarization intensity and quantum correction deformation parameters while maintaining the magnetic field configuration as $B_{(\rm r,\phi,z)}=(0.87,0.5,0)$, as depicted in Fig. \ref{fig:8}. Our findings reveal that a decrease in radius is accompanied by a gradual decrease in polarization intensity. Furthermore, the progressive increase in deformation parameters corresponds to a gradual reduction in intensity at a constant radius, indicating an anti-polarization behavior induced by quantum correction.

\section{Conclusions}
\label{sec:5}
The influence of quantum corrections on the polarized image of a black hole has been revealed in this analysis. We found that the determination of observable polarization depends on numerous physical parameters inherent to the accreting flow, including fluid velocity and magnetic field strength in the fluid's rest frame. We selected the optimal magnetic field $B_{(\rm r,\phi,z)}=(0.87,0.5,0)$ to simulate the observable polarization emanating from a stationary fluid ring. Our findings indicate that an increase in the observation inclination angle leads to a gradual flattening of the polarization image, heightened polarization intensity, and a progressive shift in polarization direction towards the southeast. Regarding the quantum correction effect, we observed that a larger deformation parameter leads to an expansion of the polarization region. This suggests that changes in the spatiotemporal structure of black holes can also impact observations.

\par
We investigated the effects of fluid direction angle and fluid velocity on polarization images. As the fluid direction angle decreases, we observed a corresponding reducing in polarization intensity, accompanied by a slight increase in the slope of the EVPA. Additionally, an increase in fluid velocity resulted in a reduction in polarization intensity without altering the potential angle of the electric vector. Despite variations in fluid direction angle and velocity, the influence of quantum correction persisted with consistent properties. This suggests a spatiotemporal feature that remains independent of external conditions. We separately considered the effects of magnetic fields, including pure radial magnetic fields, azimuthal magnetic fields, and vertical magnetic fields, on KS black hole polarization images. We found that maxima and minima in polarization intensity were obtained with the radial magnetic field. The EVPA exhibited periodic variations with azimuthal changes. Furthermore, an increase in deformation parameters corresponded to a decrease in the polarization intensity value, implying that larger deformation parameters result in diminished spatiotemporal polarization. Additionally, we observed an increase in polarization intensity on the left side of the image due to aberration, which is attributed to the motion effect inherent in any static spacetime.

\par
For the angular and radial components of the equatorial magnetic field, we observed a significant increase in polarization intensity with rising observation angles, indicating that enhancing observation inclination angles can improve the accuracy of polarization data. However, polarization intensity experiences a decline with increasing deformation parameter. As the observation tilt angle rises, the linear correlation between EVPA changes and the azimuthal period $\phi$ diminishes, resulting in non-linear distortions. As the fluid direction angles decrease, the peak polarization intensity experiences a reduction, while the alteration in EVPA remains relatively modest. The phenomenon is primarily attributed to the distortion of EVPA being predominantly influenced by magnetic fields, with the contribution from fluid direction angles being comparatively minor. When $B_{\phi}$ exceeds $B_{r}$ under zero-flow conditions, the polarization intensity exhibits an irregular curve, accompanied by quasi-periodic changes in the EVPA. Upon increasing the flow rate to $0.5$, the polarization intensity curve takes on a form resembling that of an exclusively radial magnetic field, accompanied by a concurrent decrease in the peak polarization intensity. However, the EVPA remains unchanged throughout this transition. This suggests that regardless of alterations in external conditions, the discrepancy in polarization intensity remains unaltered, with larger deformation parameters approaching $0$. Simultaneously, an increase in deformation parameters corresponds to a reduction in polarization intensity, unveiling an anti-polarization effect not previously observed in prior studies. This intricate relationship suggests that heightened tilt angles induce irregular distortions in EVPA, deviating from the patterns observed in standard Schwarzschild black holes. This outcome serves as a potential indicator to discriminate between Schwarzschild black holes and quantum corrected black holes.

\par
In the polarization vectors within the emission ring image, we observed continuous variability of polarization around the black hole. An intriguing phenomenon we noted is the contraction of the inner ring with an increase in the observer's inclination angle. This observation provides potential discriminators between Schwarzschild black holes and those subjected to quantum corrections.

\par
In our comprehensive analysis of polarization images, we computed the polarization intensity across diverse radii and deformation parameters. We established the relationship between polarization intensity and quantum correction deformation parameters, revealing a gradual decline in polarization intensity with reduced radius and an anti-polarization behavior induced by the progressive increase in deformation parameters at a constant radius. These findings suggest that the influence of quantum correction may offer theoretical evidence for the next-generation EHT to discern the type of black hole.

\acknowledgments

This work is supported by the National Natural Science Foundation of China (Grant No. 12133003, 42230207) and the Fundamental Research Funds for the Central Universities, China University of Geosciences (Wuhan) (Grant No. G1323523064).


\end{document}